\documentclass[a4paper,11pt]{article}
\pdfoutput=1 

\usepackage{jinstpub} 
\usepackage{comment}
\usepackage{graphicx}
\usepackage{subfig}

\usepackage[]{siunitx}  

\usepackage{lineno}

\title{\boldmath Deep Neural Networks for Energy and Position Reconstruction in \mbox{EXO-200}}

\author[a,1]{S.~Delaquis, \note{Deceased}}
\author[b]{M.J.~Jewell,}
\author[c,2]{I.~Ostrovskiy,\note{Corresponding author}}
\author[b]{M.~Weber,}
\author[d]{T.~Ziegler,}

\author[a,b]{J.~Dalmasson,}
\author[a,e]{L.J.~Kaufman,}
\author[c]{T.~Richards,}

\author[e]{J.B.~Albert,}
\author[d]{G.~Anton,}
\author[f,3]{I.~Badhrees, \note{Permanent position with King Abdulaziz City for Science and Technology, Riyadh, Saudi Arabia}}
\author[g]{P.S.~Barbeau,}
\author[d]{R.~Bayerlein,}
\author[h]{D.~Beck,}
\author[i]{V.~Belov,}
\author[a]{M.~Breidenbach,}
\author[j,q]{T.~Brunner,}
\author[k]{G.F.~Cao,}
\author[k]{W.R.~Cen,}
\author[l]{C.~Chambers,}
\author[m,n]{B.~Cleveland,}
\author[h]{M.~Coon,}
\author[l]{A.~Craycraft,}
\author[f]{W.~Cree,}
\author[o]{T.~Daniels,}
\author[i,4]{M.~Danilov, \note{Now at P.N.Lebedev Physical Institute of the Russian Academy of Sciences, Moscow, Russia}}
\author[e]{S.J.~Daugherty,}
\author[p]{J.~Daughhetee,}
\author[a]{J.~Davis,}
\author[m]{A.~Der~Mesrobian-Kabakian,}
\author[b]{R.~DeVoe,}
\author[q]{J.~Dilling,}
\author[i]{A.~Dolgolenko,}
\author[r]{M.J.~Dolinski,}
\author[l]{W.~Fairbank Jr.,}
\author[m]{J.~Farine,}
\author[s]{S.~Feyzbakhsh,}
\author[t]{P.~Fierlinger,}
\author[b]{D.~Fudenberg,}
\author[f,q]{R.~Gornea,}
\author[b]{G.~Gratta,}
\author[u]{C.~Hall,}
\author[r]{E.V.~Hansen,}
\author[l]{D.~Harris,}
\author[d]{J.~Hoessl,}
\author[d]{P.~Hufschmidt,}
\author[c]{M.~Hughes,}
\author[l]{A.~Iverson,}
\author[v]{A.~Jamil,}
\author[a]{A.~Johnson,}
\author[i]{A.~Karelin,}
\author[f]{T.~Koffas,}
\author[b,5]{S.~Kravitz, \note{Now at Lawrence Berkeley National Laboratory, Berkeley, California, USA}}
\author[q]{R.~Kr\"{u}cken,}
\author[i]{A.~Kuchenkov,}
\author[w]{K.S.~Kumar,}
\author[q]{Y.~Lan,}
\author[x]{D.S.~Leonard,}
\author[b]{G.S.~Li,}
\author[h]{S.~Li,}
\author[m]{C.~Licciardi,}
\author[r]{Y.H.~Lin,}
\author[p]{R.~MacLellan,}
\author[d]{T.~Michel,}
\author[a]{B.~Mong,}
\author[v]{D.~Moore,}
\author[j]{K.~Murray,}
\author[w]{O.~Njoya,}
\author[a]{A.~Odian,}
\author[c]{A.~Piepke,}
\author[s]{A.~Pocar,}
\author[q]{F.~Reti\`{e}re,}
\author[m]{A.L.~Robinson,}
\author[a]{P.C.~Rowson,}
\author[d]{S.~Schmidt,}
\author[b,6]{A.~Schubert, \note{Now at OneBridge Solutions}}
\author[f]{D.~Sinclair,}
\author[c]{A.K.~Soma,}
\author[i]{V.~Stekhanov,}
\author[s]{M.~Tarka,}
\author[l]{J.~Todd,}
\author[k]{T.~Tolba,}
\author[c]{V.~Veeraraghavan,}
\author[y]{J.-L.~Vuilleumier,}
\author[d]{M.~Wagenpfeil,}
\author[a]{A.~Waite,}
\author[f]{J.~Watkins,}
\author[k]{L.J.~Wen,}
\author[m]{U.~Wichoski,}
\author[d]{G.~Wrede,}
\author[v]{Q.~Xia,}
\author[h]{L.~Yang,}
\author[r]{Y.-R.~Yen,}
\author[i]{O.Ya.~Zeldovich}

\affiliation[a]{SLAC National Accelerator Laboratory, Menlo Park, California 94025, USA}
\affiliation[b]{Physics Department, Stanford University, Stanford, California 94305, USA}
\affiliation[c]{Department of Physics and Astronomy, University of Alabama, Tuscaloosa, Alabama 35487, USA}
\affiliation[d]{Erlangen Centre for Astroparticle Physics (ECAP), Friedrich-Alexander-University Erlangen-N\"urnberg, Erlangen 91058, Germany}
\affiliation[e]{Physics Department and CEEM, Indiana University, Bloomington, Indiana 47405, USA}

\affiliation[f]{Physics Department, Carleton University, Ottawa, Ontario K1S 5B6, Canada}
\affiliation[g]{Department of Physics, Duke University, and Triangle Universities Nuclear Laboratory (TUNL), Durham, North Carolina 27708, USA}
\affiliation[h]{Physics Department, University of Illinois, Urbana-Champaign, Illinois 61801, USA}
\affiliation[i]{Institute for Theoretical and Experimental Physics, Moscow, Russia}
\affiliation[j]{Physics Department, McGill University, Montr\'{e}al, Qu\'{e}bec  H3A 2T8, Canada}
\affiliation[k]{Institute of High Energy Physics, Beijing, China}
\affiliation[l]{Physics Department, Colorado State University, Fort Collins, Colorado 80523, USA}
\affiliation[m]{Department of Physics, Laurentian University, Sudbury, Ontario P3E 2C6, Canada}
\affiliation[n]{SNOLAB, Sudbury, Ontario P3Y 1N2, Canada}
\affiliation[o]{Department of Physics and Physical Oceanography, University of North Carolina Wilmington, Wilmington, North Carolina 28403, USA}
\affiliation[p]{Department of Physics, University of South Dakota, Vermillion, South Dakota 57069, USA}
\affiliation[q]{TRIUMF, Vancouver, British Columbia V6T 2A3, Canada}
\affiliation[r]{Department of Physics, Drexel University, Philadelphia, Pennsylvania 19104, USA}
\affiliation[s]{Amherst Center for Fundamental Interactions and Physics Department, University of Massachusetts, Amherst, MA 01003, USA}
\affiliation[t]{Technische Universit\"at M\"unchen, Physikdepartment and Excellence Cluster Universe, Garching 80805, Germany}
\affiliation[u]{Physics Department, University of Maryland, College Park, Maryland 20742, USA}
\affiliation[v]{Department of Physics, Yale University, New Haven, Connecticut 06511, USA}
\affiliation[w]{Department of Physics and Astronomy, Stony Brook University, SUNY, Stony Brook, New York 11794, USA}
\affiliation[x]{IBS Center for Underground Physics, Daejeon 34047, Korea}
\affiliation[y]{LHEP, Albert Einstein Center, University of Bern, Bern, Switzerland}

\emailAdd{iostrovskiy@ua.edu}

\abstract{
We apply deep neural networks (DNN) to data from the \mbox{EXO-200} experiment. In the studied cases, the DNN is able to reconstruct the relevant parameters -- total energy and position -- directly from raw digitized waveforms, with minimal exceptions. For the first time, the developed algorithms are evaluated on real detector calibration data. The accuracy of reconstruction either reaches or exceeds what was achieved by the conventional approaches developed by \mbox{EXO-200} over the course of the experiment. Most existing DNN approaches to event reconstruction and classification in particle physics are trained on Monte Carlo simulated events. Such algorithms are inherently limited by the accuracy of the simulation. We describe a unique approach that, in an experiment such as \mbox{EXO-200}, allows to successfully perform certain reconstruction and analysis tasks by training the network on waveforms from experimental data, either reducing or eliminating the reliance on the Monte Carlo.}


\dedicated{In memory of our dear friend and colleague Dr. S\'ebastien Delaquis.}

\begin{document}
\maketitle
\flushbottom

\section{Introduction}
\label{sec:intro}

Deep Neural Networks (DNNs) are a new class of machine learning algorithms that are gaining increasing attention in experimental high energy particle physics. DNNs are made up of multiple hierarchical neural layers, as compared to ``shallow'' neural networks, which have one hidden layer. A general introduction into DNNs can be found, for example, in Ref.~\cite{dnn_intro1}. Recent studies demonstrate the ability of DNNs to compete with or outperform more conventional approaches in detecting and localizing events in a complex detector~\cite{microboone:2017}, as well as discriminating signal events from background events~\cite{naturecomm:2014,nova:2016,next:2017,microboone:2017,pandax:2018}. In a typical application, one uses a Monte Carlo simulation (MC) to generate a large number of events for which the ``truth'' (i.e.\ the correct value of classification, or other desired quantity) is known. One then reconstructs the relevant parameters of the simulated events, such as energy and position, and trains the network with these quantities as inputs. The performance is evaluated on an independent set of MC events, not used during training. The performance of the algorithm is therefore contingent on the accuracy of both the MC and the reconstruction routines used to extract input for the network. The latter requirement already led previous investigations (Refs.~\cite{jets:2016, nova:2016, microboone:2017}) to attempt to circumvent as much of the conventional reconstruction as possible, with Ref.~\cite{microboone:2017} using almost raw waveforms (with only noise filtering and gain calibration applied) as input to DNNs. It can also be said that an algorithm trained on MC can only exploit those features of the data that were explicitly modeled in the MC.

In this work, we investigate the possibility to avoid certain conventional reconstruction steps and use DNNs to extract the energy and event position directly from the waveforms. We also consider the situation where information about a relevant parameter is accessible from distinct detection channels. This is the case in \mbox{EXO-200}, where both scintillation light and free, drifting electrons, which are produced by particles interacting in the detector, are recorded independently. We show that it is possible to use one of these two signal channels, from which the information can be more easily and accurately extracted, as a source of truth information for reconstruction of the other channel. This avoids reliance on the MC of the other channel, which in the case of \mbox{EXO-200}, can contain systematic errors leading to imperfect modeling of the data. In two out of the three considered cases, actual detector calibration data is available, and we use it to evaluate and validate the performance of the DNNs.  

The paper is organized as follows. The second section provides a brief overview of the \mbox{EXO-200} detector and the existing conventional event reconstruction chain, to the extent necessary to understand the following sections. The third section describes the DNN algorithm developed for energy reconstruction of a point charge deposited on an individual charge collection wire. A special case of induction signals on nearby collection wires is also discussed. The fourth section is dedicated to the DNN reconstruction of arbitrary charge deposits on all charge collection wires. The fifth section deals with data-driven DNN reconstruction of an event position using scintillation light in the detector. The results are summarized and discussed in the last section.

\section{Basics of \mbox{EXO-200}: Detector, reconstruction, and simulation}

\subsection{The \mbox{EXO-200} Detector}
\label{sec:detector}

The details of the \mbox{EXO-200} detector have been previously described elsewhere~\cite{Auger:2012gs}. A brief description of the detector components relevant for these studies is included here for clarity.    

The EXO-200 experiment is searching for neutrinoless double beta decay ($0\nu\beta\beta$), a theoretical nuclear decay that, if observed, would indicate that neutrinos are Majorana particles and could help constrain the absolute neutrino mass \cite{ostr2016}.  EXO-200 started taking data in Sep. 2011 and is still in operation.  The current data is divided into two periods, defined as Phase~I and Phase~II, which are separated by a hiatus due to incidents at the host site. In addition, the start of Phase~II was preceded by an electronics upgrade and a change to the electric field in the detector.  

The \mbox{EXO-200} detector consists of a cylindrical radiopure copper vessel filled with Liquid Xenon (LXe), enriched to \SI{80.672+-0.014}{\percent} in the isotope $^{136}$Xe. The detector is \SI{\sim 44}{\centi\meter} long and \SI{\sim 40}{\centi\meter} in diameter and includes two identical back-to-back time projection chambers (TPCs) that share a cathode located at the center of the detector.  

Particles interacting with the LXe deposit energy by producing both scintillation light (\SI{178}{\nano\meter}) and electron-ion pairs (ionization). Electrons are drifted from their initial location toward the anode by a uniform bulk electric field.  \mbox{EXO-200} has operated at different bulk fields ranging from \SI{20}{\volt\per\centi\meter} to \SI{615}{\volt\per\centi\meter}. For simplicity, this study only focuses on data taken during Phase~I with a drift field of \SI{380}{\volt\per\centi\meter}. ~\cite{exo200upgrade}

The charge signal (number of electrons) is detected at the anodes of each TPC by a pair of wire planes crossed at a \ang{60} angle, each of which lies in front of an array of large-area avalanche photodiodes (APDs) that detect scintillation light. The frontmost wire plane in each pair (V-Wires) serves both as a shielding grid and to detect induced signals as electrons are drifted from the interaction location to the second wire plane (U-Wires) where charge is collected. Each wire plane includes 38 readout channels with \SI{9}{\milli\meter} pitch resulting in \num{152} total charge channels. Each APD plane consists of \num{37} readout channels for a total of \num{74} APD channels (four APD channels have since been disconnected from the front-end electronics due to large leakage currents).

\begin{figure}[ht!]
\centering
\includegraphics[width = 0.8\textwidth]{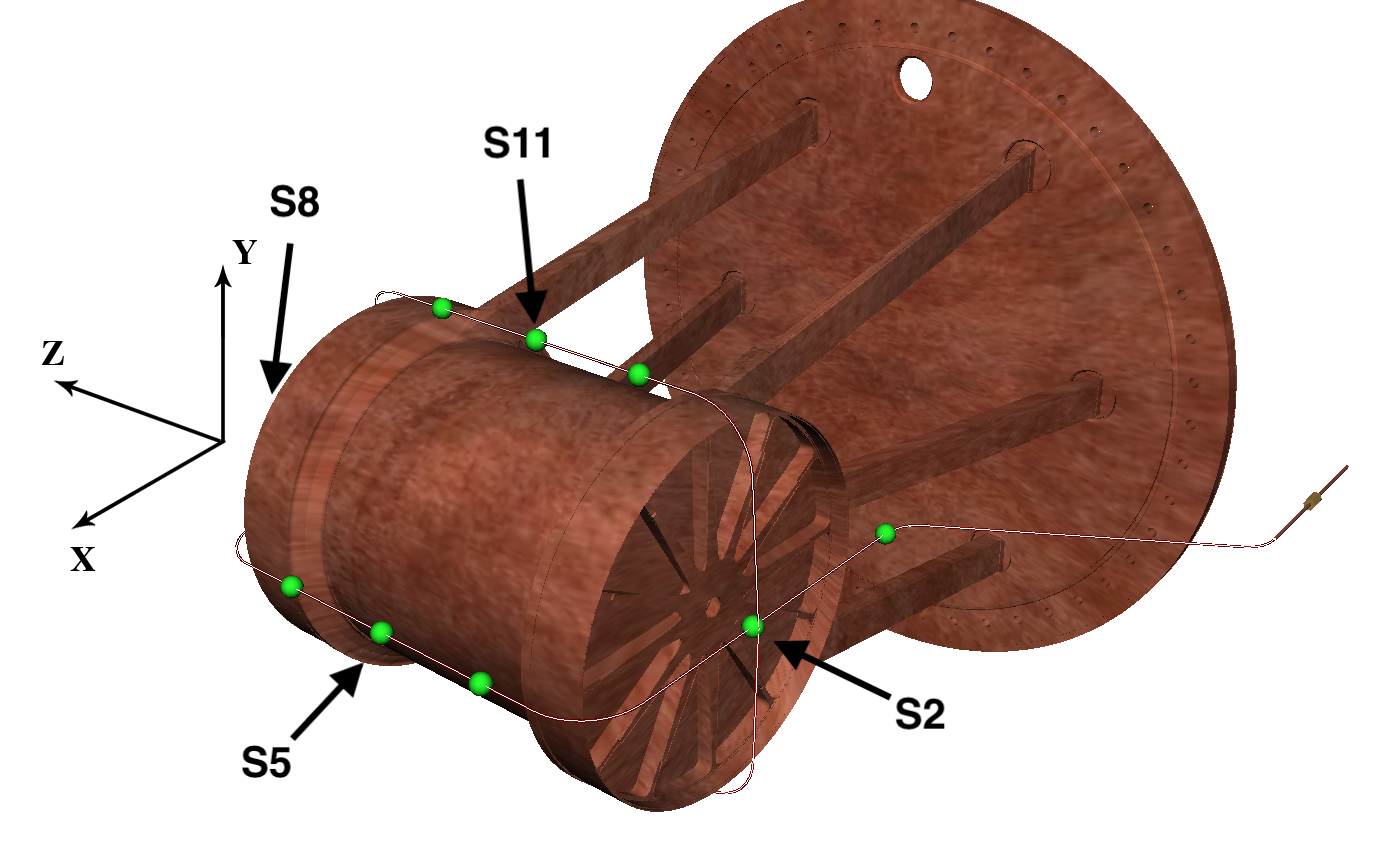}
\caption{Calibration source positions around the detector vessel. Details in the text.}
\label{fig:source_positions}
\end{figure}

Calibration of the detector is periodically performed by positioning $\gamma$-ray sources at several locations around the detector. For this publication, three sources are used ($^{60}$Co, $^{228}$Th, and $^{226}$Ra) to span the energy range of interest for the double beta decay search. These sources can be repeatedly deployed to fixed locations in a copper guide tube that wraps around the outside of the LXe TPC. The positions used for calibration (see Figure~\ref{fig:source_positions}) include two positions located around the edge of the cathode and separated by \ang{90} (named S5 and S11) and two positions centered on the drift axis behind either anode plane (named S2 and S8).

\subsection{Conventional reconstruction in a nutshell}\label{sec:EXOrec}

For each trigger in the \mbox{EXO-200} detector, the digitized raw waveforms from each of the \num{152} charge channels and \num{74} light channels are stored. Each waveform consists of \num{2048} samples taken at \SI{1}{MS\per\second} with \num{1024} samples occurring before the trigger and \num{1024} samples after the trigger. The standard analysis described in previous studies uses a conventional event reconstruction (``EXO-recon") algorithm to process and analyze the raw waveforms to determine event level information such as energy and position. The conventional reconstruction comprises three stages, which are briefly described below. A more detailed discussion of the conventional reconstruction can be found in~\cite{PhysRevC.89.015502}.

An initial signal finding stage identifies channels containing a signal above a given noise threshold. For this, a matched filter is applied to each channel using a predefined signal template to search for signals above the noise. A second iteration of signal finding is applied to the waveforms after reshaping the signals found by the matched filter. This is used to disentangle waveforms with closely-spaced signals, which the matched filter is ill-suited to find.

Following the signal finding stage is the parameter estimation stage, where each of the identified signals are analyzed to extract parameters relevant for the analysis. This is done by fitting each of the signals to their respective signal models. A $\chi^2$ fit is used to extract both the amplitude and the timing of each signal. Additionally, every charge signal is classified as either ``induction" or ``collection", which attempts to determine whether the charge deposit terminated its track on the channel. This classification is done using additional information about the waveform, such as the pulse timing, pulse integral, nearest neighbor amplitude, and the $\chi^2$ of the fit of the pulse to templates that describe the shape of induced and collected signals. A set of one dimensional (1D) cuts on these parameters is then applied to classify signals as induction or collection.

Finally, the set of signals and signal parameters for each event is combined to determine event topology and event energy. First, a ``bundling" algorithm is applied to separately group U, V, and APD channels based on their timing and amplitude characteristics. A clustering algorithm is then used to group together the bundles of different channel types into the most likely configuration based on a set of probability density functions (PDFs). This results in ``clusters," which roughly represent individual charge deposits in the detector. A cluster containing signals of all \num{3} types has a well defined 3D position and energy and is defined as a ``3D cluster". The light and charge energy of an event can be estimated from the APD and U-Wire amplitudes respectively. The Z-position comes from the timing difference between the U-Signals and the APD signals and the U- and V-position from the energy weighted combination of the U- and V-Signal amplitudes.

Following event reconstruction, corrections are applied to the reconstructed parameters to account for known effects in the detector. First, each channel is corrected for its individual gain to normalize the response of each channel to the same energy deposit. These gains are measured using calibration data, and the correction accounts for known time dependencies. Next, a correction is applied to the charge energy to remove the effects of a finite electron lifetime in the detector. This removes the Z-dependent bias introduced from electrons in LXe capturing on electronegative impurities as they drift, which attenuates the detected charge signal exponentially with the drift time. During Phase~I the APD channels experienced large time varying correlated noise that degraded energy resolution.  To account for this, a denoising algorithm was applied to the light response to optimally estimate the light energy with knowledge of the correlated noise among APD channels and the position dependent light response~\cite{Davis:2016reu}.  Electronic upgrades prior to the start of Phase~II significantly reduced this noise component and the need for denoising in Phase~II.      
Finally, in order to improve the energy reconstruction and optimize the energy resolution, a linear combination of the light and charge energy is used to calculate a "rotated" energy of an event. The rotated energy has a better energy resolution than either charge or light alone due to the anti-correlation between the light and charge signals arising from the electron-ion recombination process in LXe~\cite{conti}. 

\subsection{Monte Carlo simulation}\label{sec:MCsim}
To understand the detector response to energy deposits in the detector volume, a Monte Carlo simulation is employed. A detailed description of the simulation package is found in~\cite{PhysRevC.89.015502}, but a brief description of the details relevant to this paper are included below.   

The simulation is divided into two independent components. First, using GEANT4~\cite{G4Paper,Agostinelli2003250}, the \mbox{EXO-200} detector geometry is modeled and the physical interactions in the detector are implemented. GEANT4 returns a list of charge depositions within the detector volume from the relevant ionizing  interactions. These depositions are sampled with cubic voxels with \SI{0.2}{\micro\meter} edges.    

The set of depositions is then passed into the second component of the simulation, where the electronic response of the system is modeled to produce waveforms similar to those recorded during data taking. For charge waveforms on the U/V planes, the signal generation step tracks each charge deposit through the detector using a 3D finite element simulation of the electric field to realistically model the path of the electrons. Diffusion is also included to correctly reproduce the observed channel multiplicity~\cite{EXODiff}. The effects of a finite electron lifetime are also optionally incorporated to model the Z-dependent response. At every time step during the drift, the charge induced on each readout channel by the drifting electrons is determined from the Shockley-Ramo theorem~\cite{Shock,Ramo} which produces a signal in each affected channel. The signals are then shaped using the known transfer functions of the readout electronics for each channel and then digitized. To model the APD response, the current simulation uses a parameterized light response function to give the light yield on each APD plane based on the position of the energy deposit in the detector. The light response is then evenly distributed among all channels on the given plane and assumed to be a step function. The un-shaped signal is then transformed with the known transfer function. This does not include the effect of recombination fluctuations between the light and charge yield in LXe.
\begin{figure}[h!]
\centering
\includegraphics[width = 0.8 \textwidth]{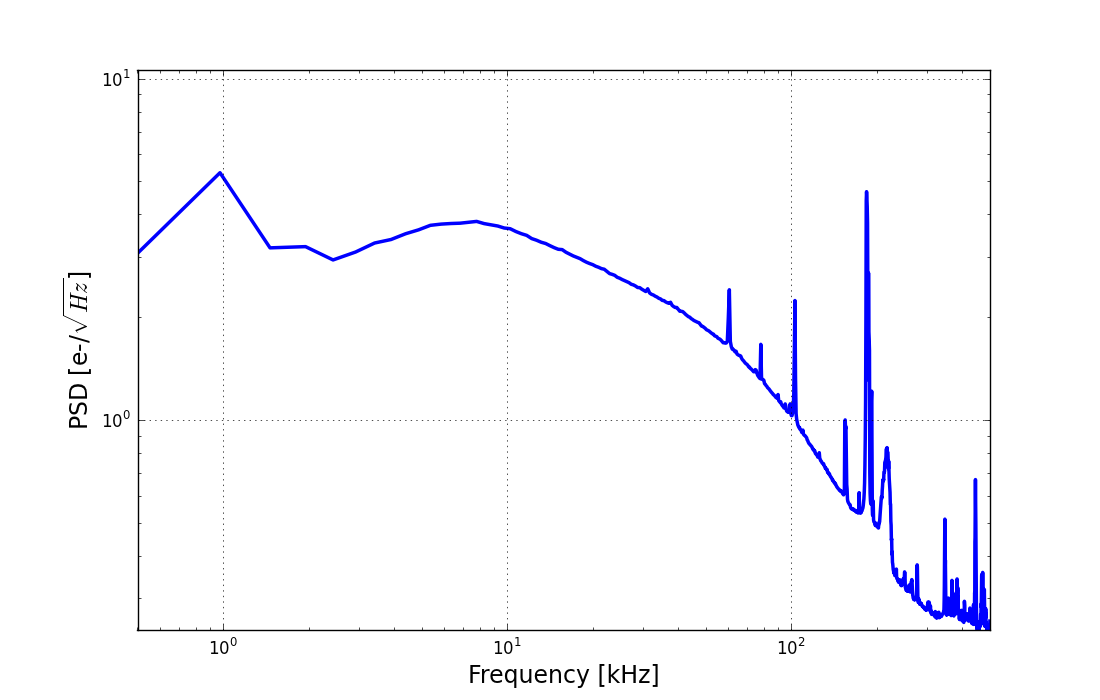}
\caption{Average power spectral density (PSD) of U-Wire noise traces in the noise library used for adding noise to the waveforms of the MC simulation.}
\label{fig:avg-fft}
\end{figure}
For this reason, the light simulation is used only to approximately simulate the light reconstruction threshold and is not used to determine the Monte Carlo energy. 

Finally, in order to make simulated waveforms more realistic, randomized noise waveforms recorded periodically during data taking are added to each of the simulated waveforms on both the charge and light channels. This creates MC waveforms for each channel with a noise spectrum  similar to the true spectrum observed in data. The absolute noise varies for each channel, but on average the broadband noise observed on the U-Wire waveforms is $\sim$800~electrons (which at \SI{380}{\volt\per\centi\meter} corresponds to \SI{\sim 16}{\kilo\electronvolt}). Figure~\ref{fig:avg-fft} shows the average power spectral density of U-Wire noise traces.

\section{Energy reconstruction of point-like charge deposition events}


The energy resolution of \mbox{EXO-200} is fundamental for separating potential $0\nu\beta\beta$ decay from background events. Reconstructing the deposited energy from the waveforms of the charge and light channels is limited by an irreducible error due to noise and statistical fluctuations. In the case of the charge channels, the induction signals on U-Wires can also contribute to the reconstruction errors, if not identified and corrected for. The irreducible error sets the limit to the best possible data analysis. The energy reconstruction from a single waveform is directly linked to this error and is thus a good estimator for possible improvements in an existing data analysis. In this study, we present a method of reconstructing the energy of a point-like charge deposition from the waveform of a single U-Wire using a DNN.

This study is split in three steps. First, a DNN is trained with MC data. Second, the accuracy of charge collection reconstruction is evaluated and the DNN performance is compared with the \mbox{EXO-200} energy reconstruction algorithm (EXO-recon) on a different set from the same MC distribution. Lastly, the fraction of mis-reconstructed induction events is compared between the DNN and the EXO-recon.

\subsection{Network architecture and training on Monte Carlo data}

\begin{figure}[h]
\centering 
\includegraphics[width = 0.8\textwidth]{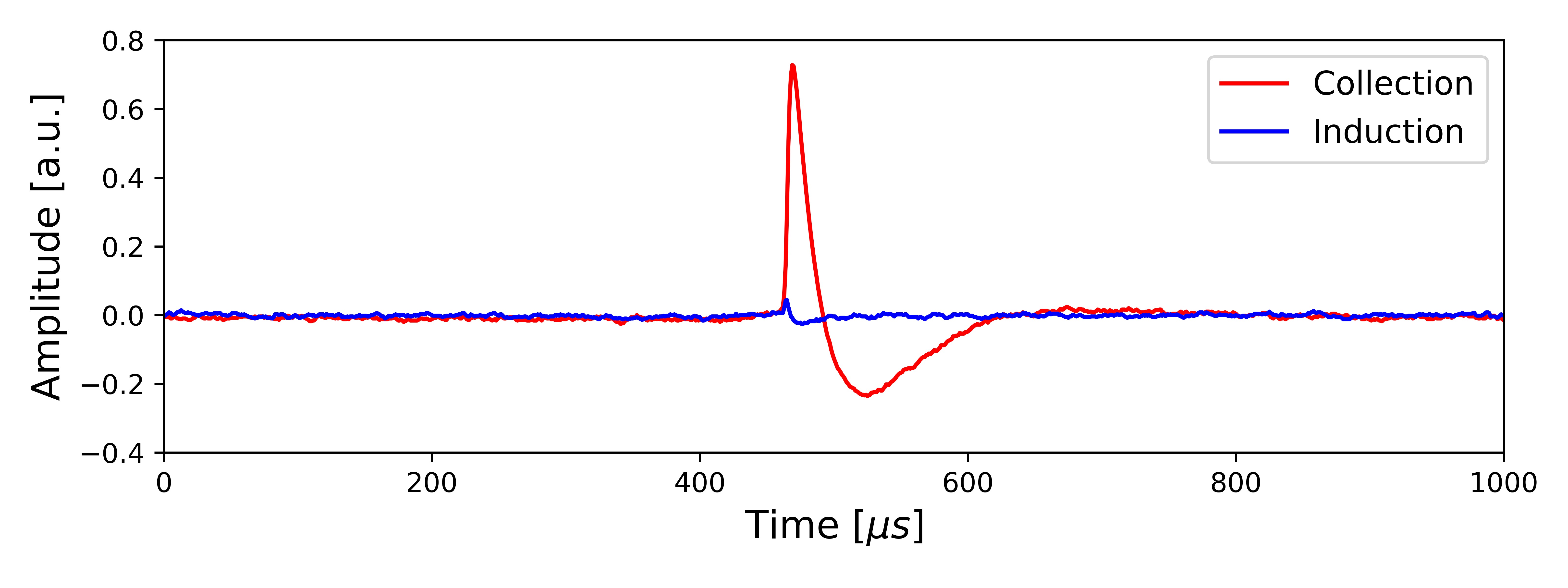}
\caption{The plot shows two waveforms of the Monte Carlo data; in red a collection waveform and in blue an induction waveform.}
\label{fig:U-WF_MC}
\end{figure}

We have built a DNN that takes a waveform as input and predicts the energy of the charge deposition. To train the DNN, we used a supervised learning method. Supervised learning requires feeding not only the input to the DNN, but also comparing the network's output to the true value. However, the true energy of a waveform collected with \mbox{EXO-200} is unknown \textit{a priori}. Therefore, we used MC, where the true energy is known, to simulate waveforms on a single wire channel. As described in Section~\ref{sec:MCsim}, the MC waveforms contain noise samples from actual \mbox{EXO-200} data. The MC set used in this study contains \num{\sim 3e5} waveforms generated from point like charge deposits artificially generated with the detector MC. One third of the waveforms represent charge collection events on a single U-Wire where the point-like deposit terminates its track. These events have their energy distributed uniformly between \SI{0}{\mega\electronvolt} and \SI{3}{\mega\electronvolt}. The remaining waveforms represent induction signals on the two neighboring U-Wires, which do not collect charge but sense an induced signal produced by the drifting charge before it terminates its trajectory. Although some charge is induced on every U-Wire channel on the wire plane, the two neighboring channels are the most likely to have a signal above the noise due to their proximity to the collecting channel. These events are marked as induction signals and have zero true energy associated with them. Examples of charge collection and induction waveforms are shown in Figure~\ref{fig:U-WF_MC}. While a full \mbox{EXO-200} waveform contains 2048 time entries, only the central 1000 entries -- 500 before and after the trigger -- are retained for this study. This reduces the computational load while keeping sufficient data for the baseline and noise estimation.

When training the DNN, we split the data into three independent sets. First, the training set, containing \SI{80}{\percent} of all MC events, is used to train multiple different DNNs. Second, a validation set (\SI{10}{\percent} of all MC), is used to select the best DNN, based on its performance over this set. Third, a test set (\SI{10}{\percent} of all MC) is used for evaluating the performance of the selected DNN. This procedure guarantees that the DNN doesn't overfit either the training set or the validation set. Overfitting is defined here as the production of an analysis which corresponds too closely or exactly to a particular set of data, and may therefore fail to fit additional data or predict future observations reliably~\cite{overfit_wiki}.

\begin{figure}[t!]
\centering
\includegraphics[width = 0.9\textwidth]{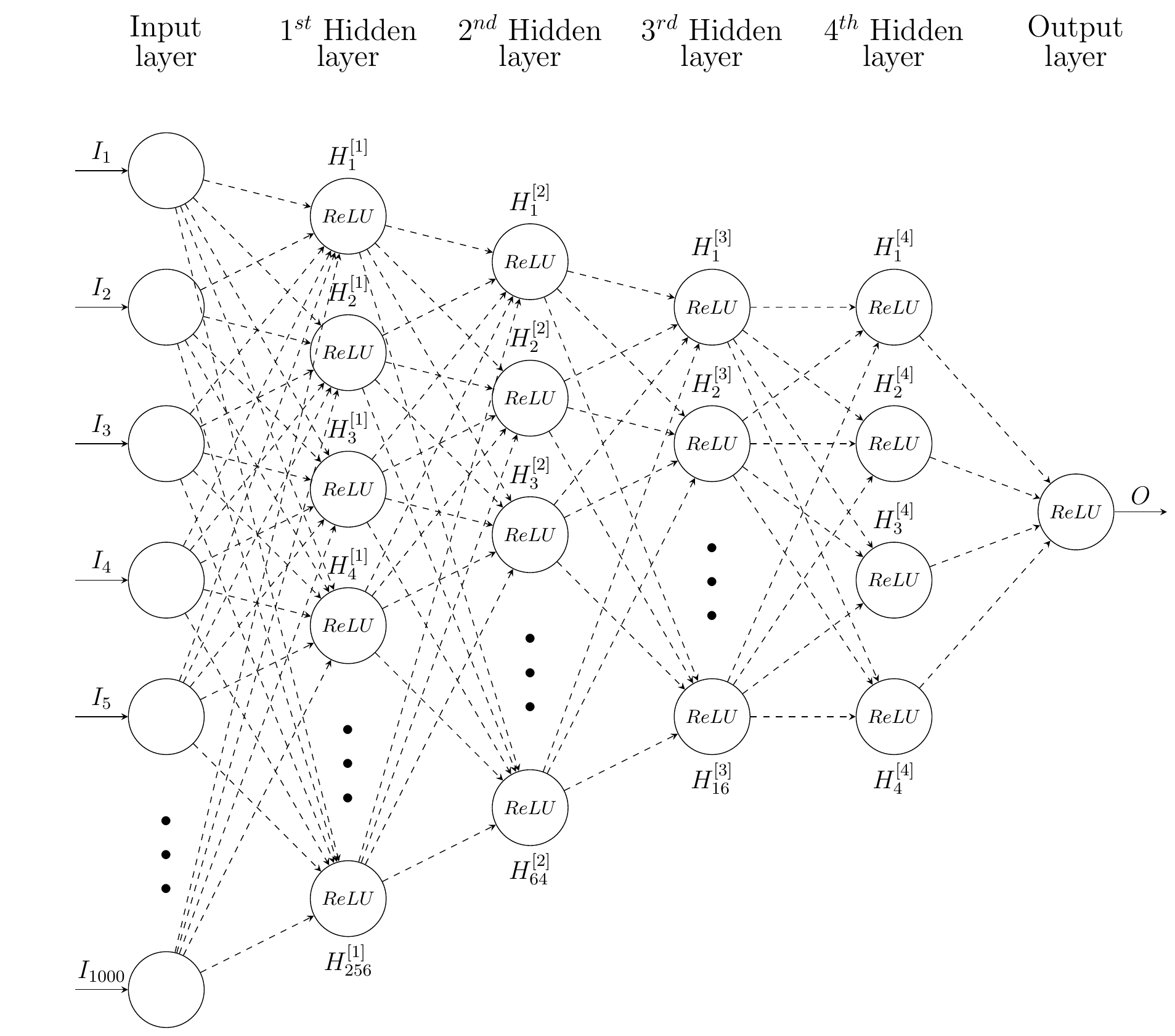}
\caption{The architecture of the DNN used for the U-Wire study. The input layer represents the input waveform with \num{1000} data points. It is fed to the first hidden layer, which has \num{256} neurons with rectified linear unit (ReLU) activation functions. The number of neurons per layer is gradually decreased from \num{256} to \num{64}, then to \num{16}, then to \num{4}, and then to \num{1} in the output layer. The DNN is trained to output the true energy in the output layer. The model contains \num{273817} trainable parameters.}
\label{fig:U-arch}
\end{figure}

A DNN can be parametrized by many parameters such as the number of hidden layers, number of neurons per layer, activation functions, layer type, and regularization techniques. To our knowledge, there are no universal algorithms for choosing an optimal network architecture. Typically, the number of neurons in the hidden layers is chosen to be less than in the input and more than in the output layers.  We trained several different network architectures containing only fully connected layers and chose the one that gave the best performance. The architecture of the DNN with the best performance is shown in Figure~\ref{fig:U-arch}. It is built from \num{5} fully connected layers. A neuron $i$ in a layer $l$ connects with a certain weight vector w$^{[l]}_{[i]}$ to all neurons in the following layer. The number of neurons per layer gradually decreases from \num{1000} (the number of samples per waveform) in the input layer, to \num{1} in the output layer, a scalar predicting the energy of the charge deposition. From layer to layer, the number of neurons decreases by a factor of about \num{4} to \num{256}, then to \num{64}, then to \num{16}, then to \num{4}, and then to \num{1}. Except for the input layer, all layers use the rectifier linear unit (ReLU). This means that an output of a neuron in those layers is given by the following function of its input:
\begin{equation}
ReLU(x) = max(0,x)
\label{eq:relu}
\end{equation}
where $x$ is the weighted sum of the outputs of all neurons from a preceding layer that connect to this neuron. ReLU is currently the most commonly used activation function~\cite{dnn_intro1}. 

To train the network we use mini-batch gradient descent to minimize the loss function $L$. In this approach, the gradient of the loss is computed using a subset of the training data, called a mini-batch, which is chosen randomly during each minimization step. Therefore, it takes more than one step to go through all the training data; the corresponding number of steps is called an epoch. We use an Adaptive Moment Estimation algorithm, or Adam~\cite{DBLP:journals/corr/KingmaB14}, to compute parameter updates at each step. The loss is defined as $L = C + \lambda \cdot R$, where $C$ is the mean square error, and $R$ is the regularization term. The mean square error is defined as:

\begin{equation}
C = \frac{1}{m} \sum_{j=1}^{m}  {\left( y_{j} - \widehat{y_{j}} \right)}^{2}
\label{eq:U-loss}
\end{equation}

\noindent{}where $m$, $ y_{j}$, and $\widehat{y_{j}}$ are the size of the mini-batch, the true energy, and the predicted energy of event~$j$, respectively. The regularization parameter $\lambda = \num{e-3}$ defines the contribution of the regularization term ($R$) to the loss ($L$). Regularization was implemented using weight decay, also known as $L$2:

\begin{equation}
R = \frac{1}{N} \sum_{l}^{Lr} \sum_{i}^{N_l}{\left(w_{[i]}^{[l]}\right)}^{2}
\label{eq:U-reg}
\end{equation}

\noindent{}where $N$, $Lr$, $N_l$, and $w_{[i]}^{[l]}$ are the number of weights in the entire DNN, the number of layers, the number of weights in the $l^{th}$ layer, and the weight vector $i$ of layer $l$, respectively. The network is set up and trained using the TensorFlow framework~\cite{tensorflow}. The framework contains implementation of mini-batch gradient descent, Adam optimizer, and other relevant algorithms.  

\begin{figure}[h]
\centering
\includegraphics[width = 0.75 \textwidth]{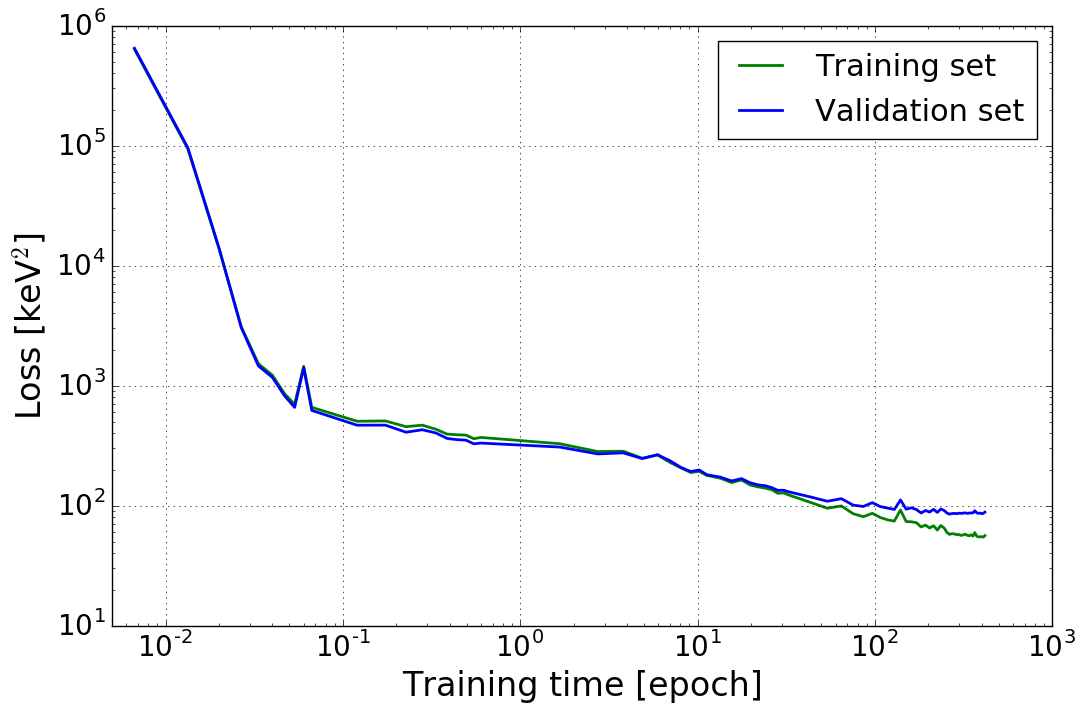}
\caption{The learning curve shows the decreasing loss during training of the DNN. Thus, the DNN is learning to predict the energy from a waveform.
}
\label{fig:U-LC}
\end{figure}

Training was done in several steps, gradually increasing the mini-batch sizes and the number of training steps per mini-batch size. During the training process, the loss was computed for the training and validation sets (see Figure~\ref{fig:U-LC}). The separation between the performance on the validation set and the training set, occurring at a training time of about \num{e2} epochs, is a sign of overfitting. However, the loss on the validation set remains stable after $\num{2}\times\num{e2}$ training epochs. To reduce overfitting, we tried to increase $\lambda$. However, with $\lambda > \num{e-3}$, $L$ was dominated by $R$, and the performance on $C$ was penalized. To reduce the overfitting, substantially increasing the size of the training set could help, which was not done in this study.

\begin{figure}[h]
\centering
\includegraphics[width = 0.9 \textwidth]{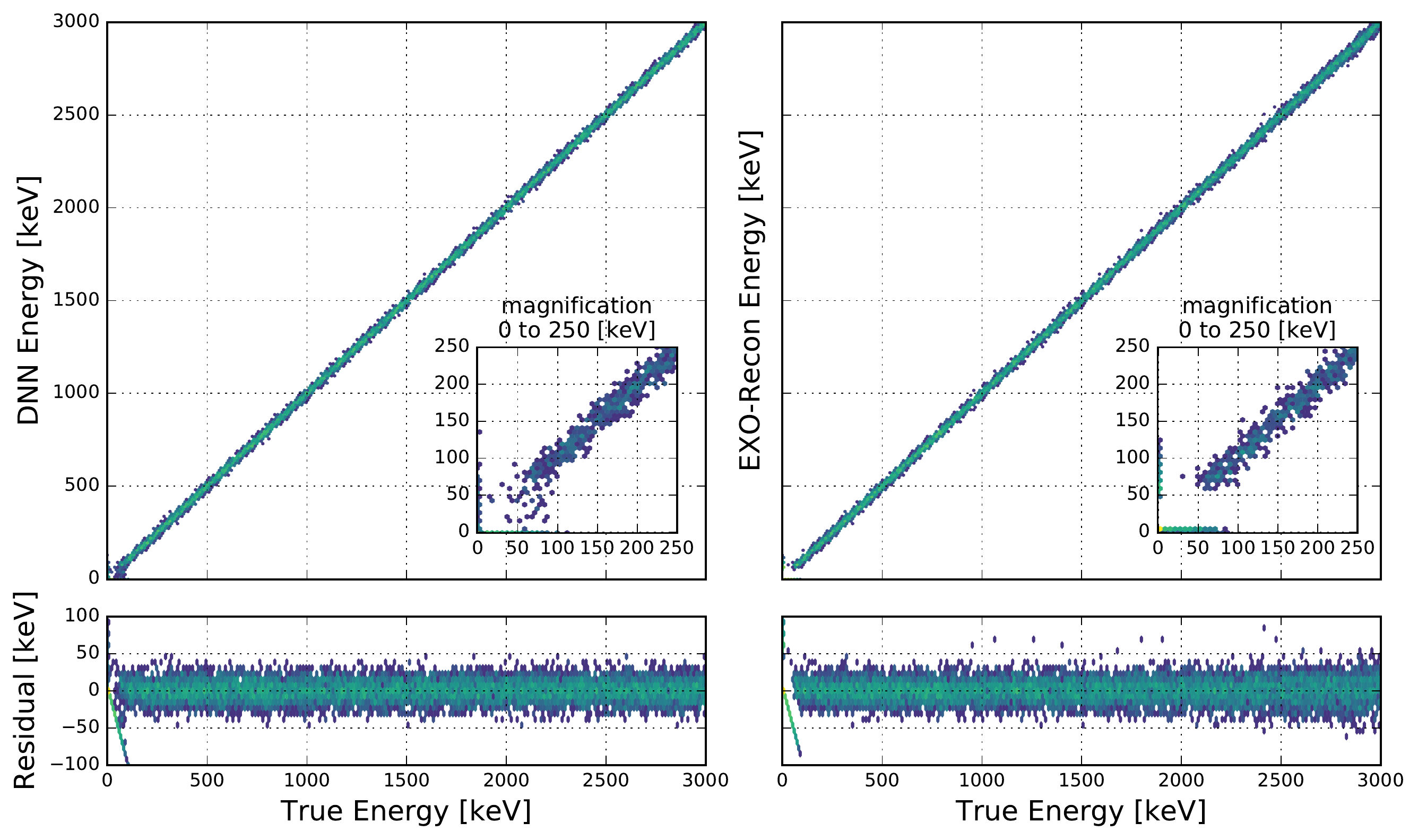}
\caption{Predicted/reconstructed energy for the test set with both methods, DNN and EXO-recon.}
\label{fig:U-pred}
\end{figure}

\subsection{Testing the network on Monte Carlo data}
\begin{figure}[h]
\centering 
\includegraphics[width = 0.7 \textwidth]{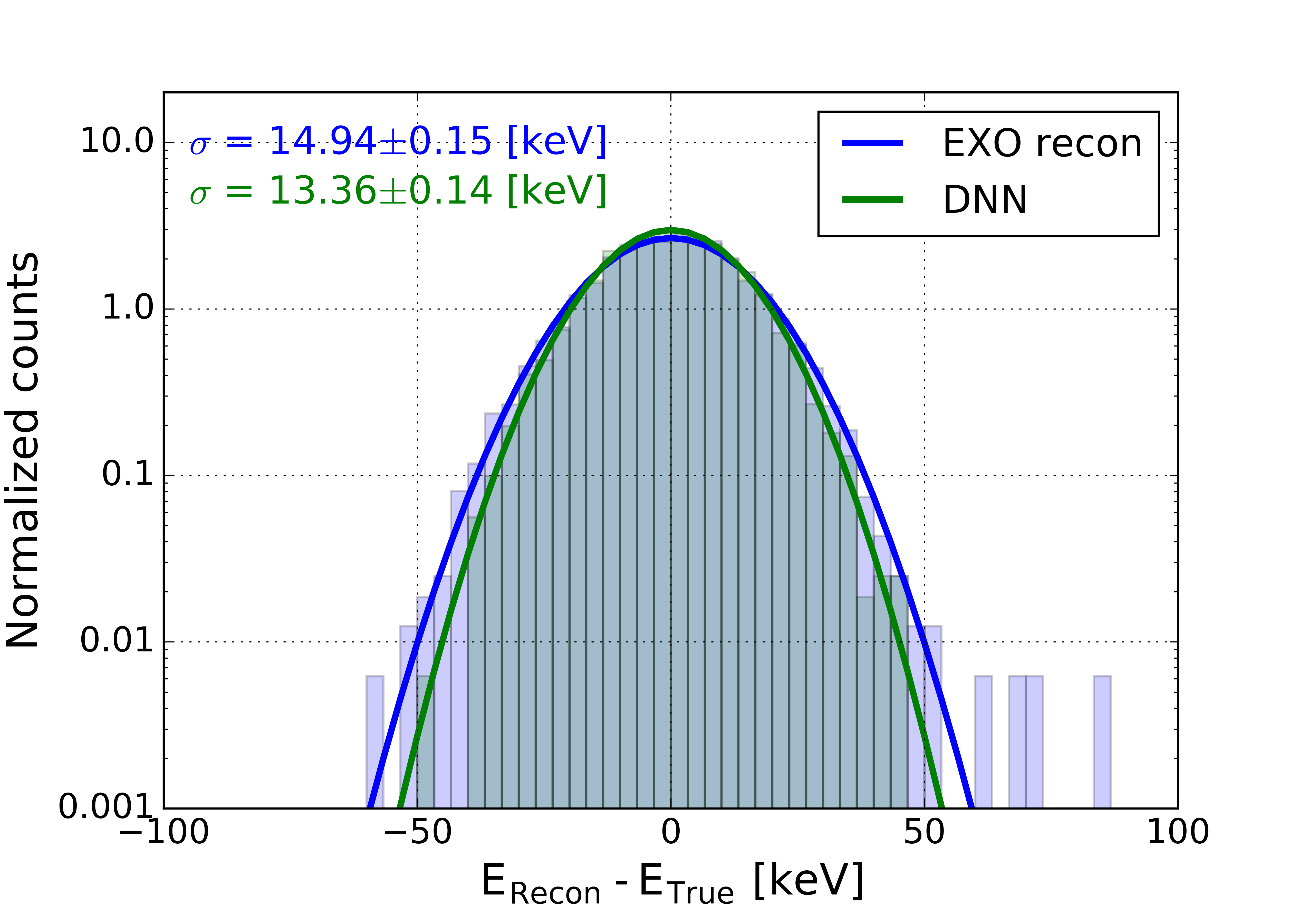}
\caption{Distribution of the residuals, predicted/reconstructed energy minus true energy, for both methods. Events with zero true energy (the induction events) are not included.}
\label{fig:U-comp}
\end{figure}

With the trained DNN, the energies of the waveforms in the test set were predicted. Figure~\ref{fig:U-pred} shows the true energy versus the predicted energy of both the DNN and EXO-recon. For energies below \SI{\sim 50}{\kilo\electronvolt} (see zoom), the DNN can still partially predict the energy while EXO-recon begins to fail. Furthermore, the residuals of the energy predicted by the DNN show fewer outlier points. The distribution of the residuals (see Figure~\ref{fig:U-comp}) is \SI{10}{\percent} narrower for the DNN compared to EXO-recon. Hence, we conclude that the DNN performs slightly better on the MC test set than EXO-recon does. The width of the distributions closely corresponds to the average U-Wire noise (see Section~\ref{sec:MCsim}), indicating that both algorithms approached a limit of performance.

\subsection{Induction on collection wires}

Upon further inspection of the reconstruction performance, a small fraction of events was found that have a zero true energy but a non-zero reconstructed energy. These are the cases when an induction signal is mistakenly reconstructed as a charge collection. In earlier \mbox{EXO-200} analyses, mis-reconstructed induction signals led to a noticeable reduction of reconstruction efficiency~\cite{exo:2012}. The conventional reconstruction algorithm was then adapted to include a dedicated procedure to discriminate induction signals.  This is done by applying a set of 1D cuts to parameters extracted from the found U-Wire signals, as mentioned in Section~\ref{sec:EXOrec}.        

\begin{figure}[htp!]
\centering 
\includegraphics[width = 0.6 \textwidth]{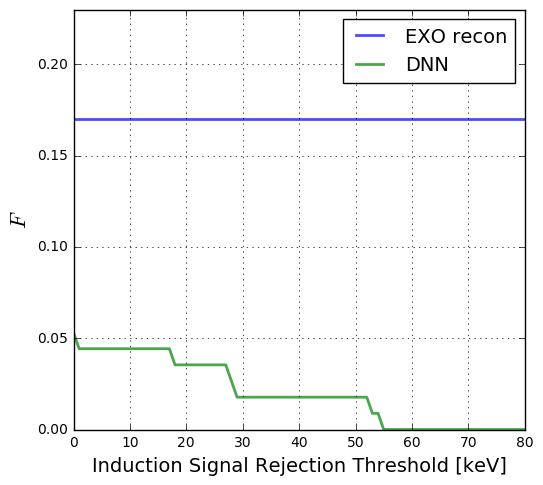}
\caption{Fraction, $F$, of induction signals that were assigned a non-zero reconstructed energy by the EXO-recon and by the DNN. For the DNN case, the fraction is shown as a function of the energy cut.}
\label{fig:ind}
\end{figure}

The test MC set contains \num{\sim 3e4} simulated signals, with each collection signal accompanied by two induction signals on the neighboring U-Wires. In this MC set, EXO-recon initially identifies \num{113} collection signals on the set of waveforms associated with true induction signals. Of these \num{113} found signals \num{94} are correctly classified by the induction classification step described above. The remaining \num{19} signals remain mis-classified as collection signals as they fail at least one of the included cuts. This gives a mis-reconstruction rate of \SI{\sim 17}{\percent}, where the mis-reconstruction rate is defined as the number of true induction signals mis-classified as collection, divided by the total number of signals initially found by EXO-recon on induction channels. The rate determined here agrees within the statistical uncertainty with the previously published value of \SI{\sim 23}{\percent}~\cite{PhysRevC.89.015502}. 

We considered developing a standalone DNN specifically for binary classification of induction and collection signals prior to energy reconstruction. However, the DNN described in this section assigns very low values of reconstructed energy to induction signals, allowing one to discriminate them with an induction signal rejection threshold. The threshold could be set to zero or another value well below the average noise level (Section~\ref{sec:EXOrec}) and still perform competitively\footnote{This does not mean that a given induction signal was below the noise threshold. Rather, a detectable induction signal was recognized and assigned energy very close to zero (true value).}. This implies that the network learned to sufficiently recognize induction signals as a by-product of the main task. To perform a direct comparison with the EXO-recon, we applied the DNN to the \num{113} (\num{94} + \num{19}) events mentioned above. We find that six of these events are assigned a non-zero reconstructed energy. Figure~\ref{fig:ind} shows the fraction of mis-reconstructed induction signals as a function of the threshold position for the DNN approach. The EXO-recon fraction is also shown for comparison. EXO-recon does not apply an energy cut to identify induction signals, so the fraction is not a function of threshold. The DNN approach slightly outperforms that of EXO-recon.  This check only focused on signals which EXO-recon found and mis-classified initially.  In addition there are in total 41 signals which are mis-classified by the DNN but not included in the above subset.  Of these signals, 21 were found by EXO-recon during the induction matched filter stage of reconstruction and were correctly classified as induction. The remaining 20 signals are under the EXO-recon threshold and are not identified as either collection or induction.  Since the DNN does not distinguish between induction and noise a direct comparison of these signals is difficult. To fully investigate the potential of the DNN to differentiate induction and collection signals one needs to create a neural network specifically for signal classification, which is not done in this study. Additionally, it is important to note that both algorithms use a simulated shape and amplitude distribution of induction signals. The accuracy of these algorithms when applied to experimental data will depend on the precision of the simulation, and any mis-modelings may result in a higher mis-reconstruction rate than seen in MC.

\section{Charge-only energy reconstruction}
\subsection{Network topology and training}\label{sec:FullE_Train}
\begin{figure}[h!]
\centering 
\includegraphics[width=\textwidth]{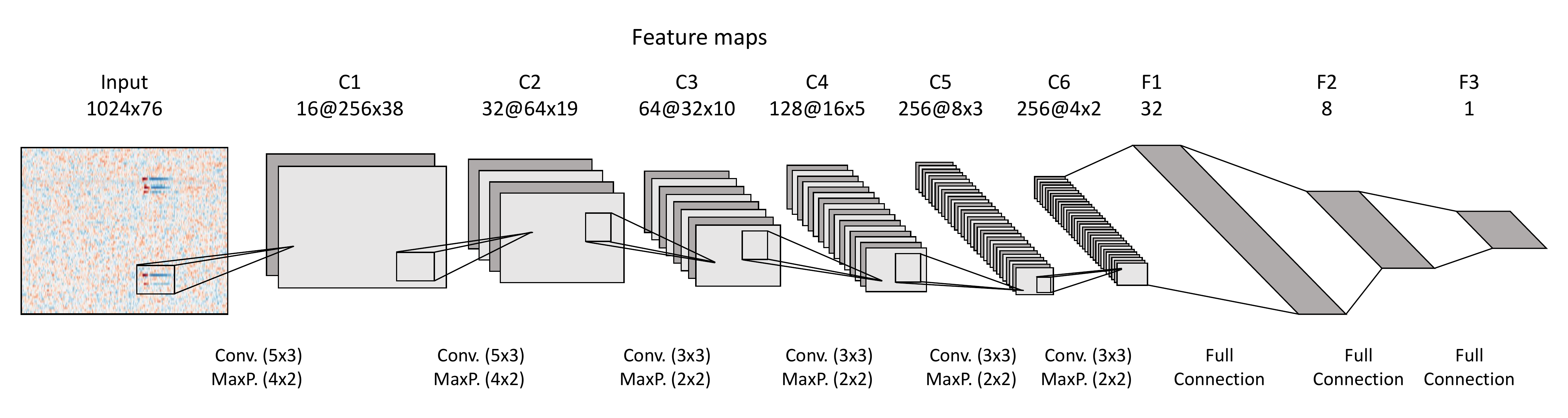}
\caption{Depiction of the network topology used for the charge-only energy reconstruction. The input image has dimensions of $\num{1024}\times\num{76}$ pixels that are fed to the network. The convolutional part consists of alternating layers of convolutions~(Conv.) and Maximum pooling~(MaxP.) resulting in feature maps with particular dimensions as given in the figure. The dimensions of the feature maps decrease after each convolutional layer while the number of kernels increases. The final feature maps are transformed into a one-dimensional array and are fed to fully connected layers where the last layer has a single unit corresponding to the deposited energy in that image.}
\label{fig:FullE_Network} 
\end{figure}

In this study, we investigate the energy reconstruction of events in the detector using all available charge collection wires.
To do so, we combine the waveforms of all individual U-Wire channels to form an image. The pixel values of the resulting image correspond to the waveforms' amplitudes at a particular time. The study is performed with events regardless of the number of charge deposits and the multiplicity of the events. With the multiplicity, one can classify the events as single-site (SS) or multi-site (MS) events. Preprocessing consists of subtracting a channel-dependent baseline obtained by averaging the first \SI{800}{\micro\second} and applying a channel-dependent gain correction.
\begin{figure}[htp]
\centering 
\includegraphics[width=0.9\textwidth]{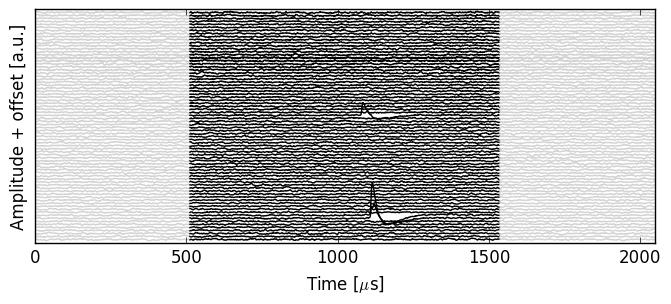}
\caption{Example waveforms on the collection wires showing multiple deposits in the detector. For clarity, the signals on each wire are offset vertically. Due to computational reasons, the first and last \num{512} time bins (light gray) are omitted, when being fed to the neural network.}
\label{fig:FullE_InputWvf}
\end{figure}
In order to save computational resources, the waveform's first and last \num{512} entries are cropped. The resulting image of the event consists of \num{76} channels, each having \num{1024} time entries, is shown in Figure~\ref{fig:FullE_InputWvf}. The image is fed to a deep convolutional neural network consisting of six successive convolutional layers. The first convolutional layer consists of 16 kernels of size $\num{5}\,\text{(time-wise)}\times\num{3}\,\text{(channel-wise)}$ pixels. Kernel corresponds here to a neuron that at any moment is only connected to a part of the input image, called receptive field. The size of the receptive field is defined by the kernel's size. Each kernel is convolved (cross-correlated) with the input image. The result of the convolution is stored as a two-dimensional array called a feature map. The 16 feature maps from the first convolutional layer are fed as input to the second convolutional layer, which has 32 kernels of the same size as in the first convolutional layer. The following convolutional layers have kernel sizes of $\num{3}\,\text{(time-wise)}\times\num{3}\,\text{(channel-wise)}$ pixels. The number of kernels gradually increases to \num{64}, \num{128}, and \num{256} in the last two layers. After the first two convolutional layers, the dimensions of the feature maps are decreased by maximum pooling by a factor of $\num{4}\,\text{(time-wise)}$ and a factor of $\num{2}\,\text{(channel-wise)}$. This means that an output of one layer is divided into a set of non-overlapping rectangles, and for each such rectangle only the maximum pixel value is retained to produce a smaller input image for the next layer. Maximum pooling by a factor of \num{2} in both dimensions is used for the subsequent layers. The network's topology is shown in Figure~\ref{fig:FullE_Network}. 
Zero padding, or surrounding the two-dimensional input image with zeros, is applied to each layer. This allows the output to have the same dimension as the input. Following the convolutional layers are three fully connected layers consisting of \num{32}, \num{8}, and a final unit. The final unit represents the deposited energy of the image corresponding to that event. Throughout the network, the ReLU activation function is used. Initial values of trainable parameters are set using the Glorot algorithm~\cite{pmlr-v9-glorot10a}. The network is implemented and trained using the Keras framework~\cite{chollet2015keras}. The framework contains implementation of all relevant algorithms. 

\begin{figure}[htp!]
\centering 
\includegraphics[width=0.6\textwidth]{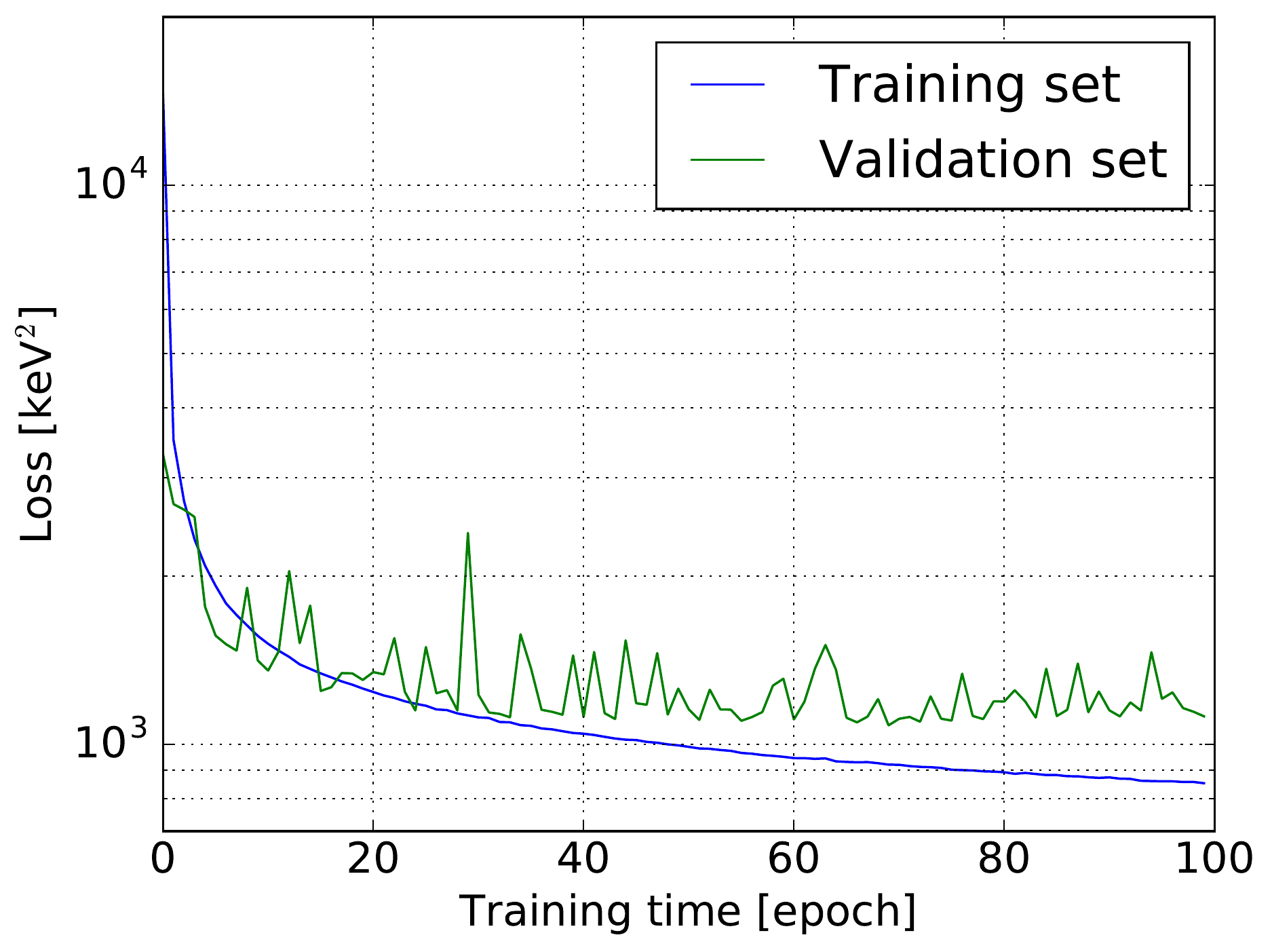}
\caption{The learning curve shows a decreasing loss during training for both the training set and the validation set. Thus, the convolutional neural network is learning to predict the energy deposited on the charge collection wires.}
\label{fig:FullE_Loss}
\end{figure}

The neural network is trained with the Adam optimizer by minimizing the mean square error. An additional $L2$ regularization term weighted with \num{e-2} is applied in each layer. MC events are generated using the standard \mbox{EXO-200} MC framework discussed in Section~\ref{sec:MCsim}. For training and validating the neural network, MC events from a gamma ray source positioned in the center of the detector are used. The gamma ray source produces an energy deposition that is uniform in energy. This is motivated by the initial but unsuccessful approach utilizing a $^{228}$Th MC source for training, see Section~\ref{sec:FullE-Testing}. The noise on the wire channels is sampled from a real noise library (see Section~\ref{sec:MCsim}). MC events that are not fully 3D reconstructed within the standard \mbox{EXO-200} framework are rejected. Additionally, events that have charge deposits outside the fiducial volume~\cite{PhysRevC.89.015502} are rejected. The true energy is defined as the sum of all pixelated charge deposits in the MC. The network is trained on events with a total deposited energy larger than \SI{500}{\kilo\electronvolt}. For training and validation, about \num{750000} MC events from a flat energy distribution in batches of \num{16} samples are used, with \SI{95}{\percent} used for training and the remaining \SI{5}{\percent} for validation. Figure~\ref{fig:FullE_Loss} shows the learning curve of the network over \num{100} epochs.

\subsection{Testing on Monte Carlo data}\label{sec:FullE-Testing}
\begin{figure}[ht!] 
    \centering
    \subfloat[DNN]{\includegraphics[width=0.48\textwidth]{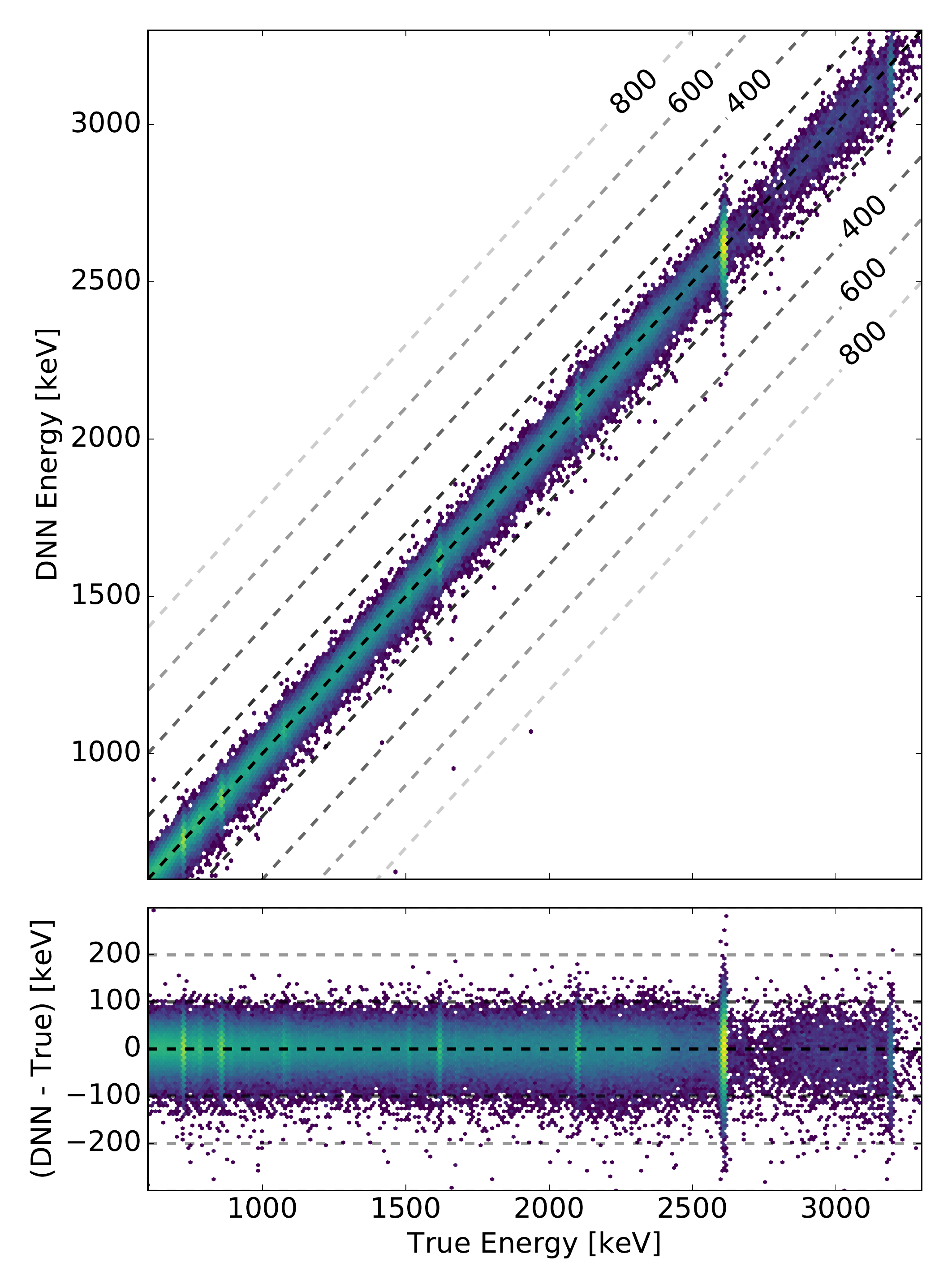}\label{fig:FullE_Prediction_train_CNN}}
    \quad
    \subfloat[EXO-recon]{\includegraphics[width=0.48\textwidth]{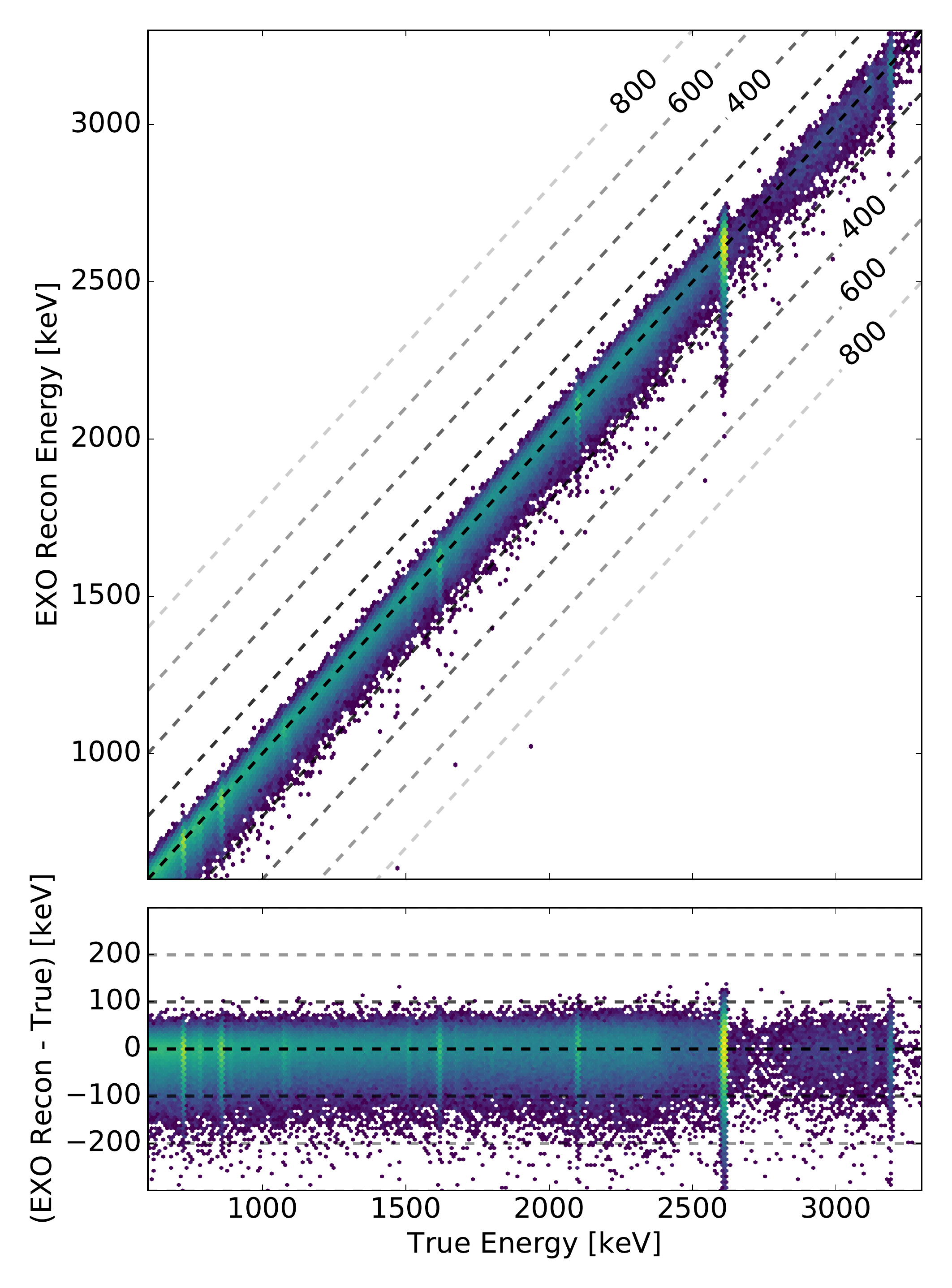}\label{fig:FullE_Prediction_train_Std}}
    \caption{Reconstructed energy of the neural network~\textbf{(a)} and the conventional analysis~\textbf{(b)} as function of the true MC energy. The lower panels show the corresponding residuals. The color denotes the intensity on a logarithmic scale. Dashed lines indicate a certain residual value as given in the figure.}
    \label{fig:FullE_Prediction_train} 
\end{figure}

\begin{figure}[ht!]
\centering 
\includegraphics[width = \textwidth]{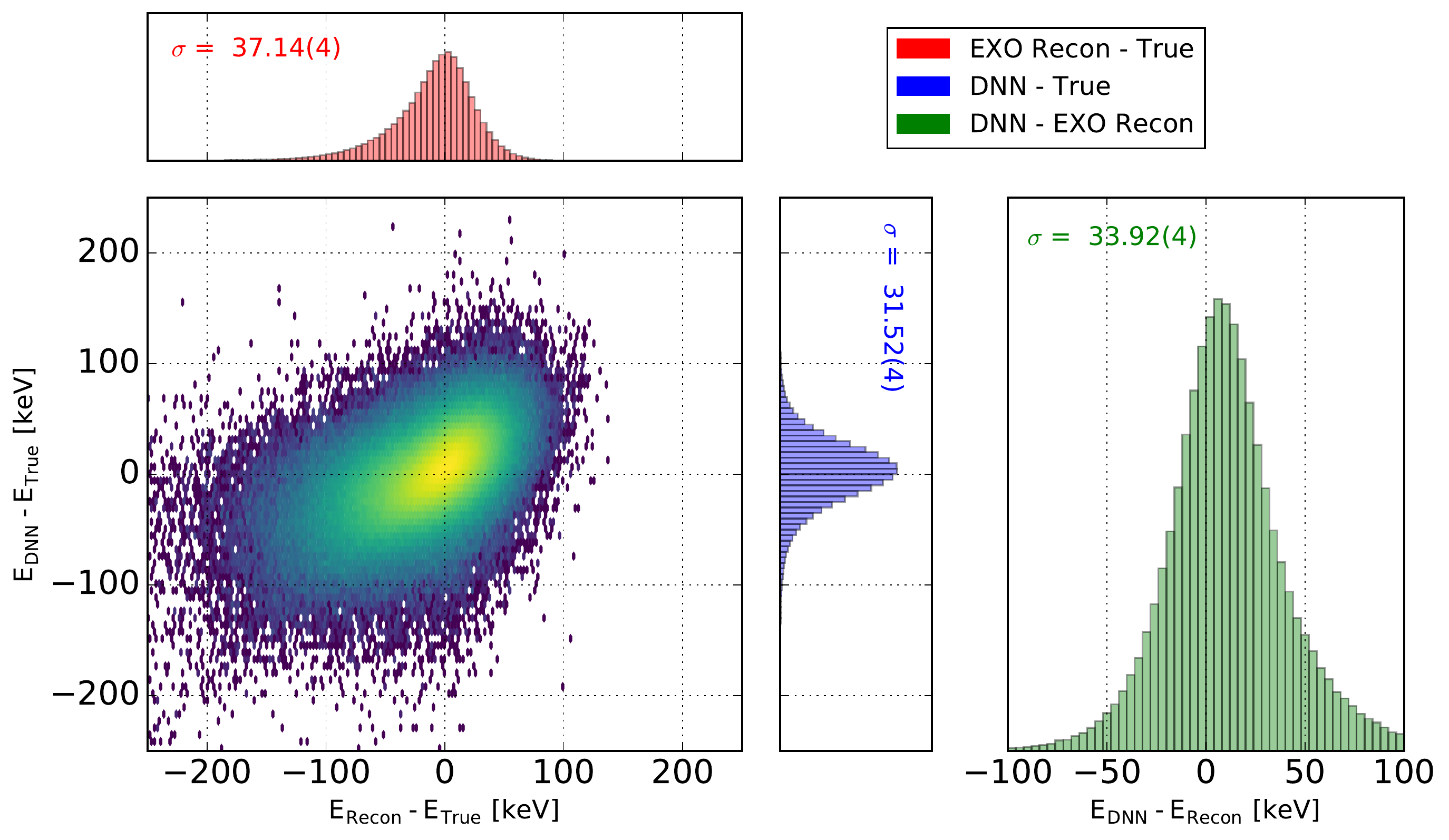}
\caption{The two dimensional distribution shows the residuals of the energy from the conventional reconstruction, $E\textsubscript{Recon}$, and from the neural network, $E\textsubscript{DNN}$, with respect to the true MC energy $E\textsubscript{True}$. The intensity is denoted by color on a logarithmic scale. The positive correlation indicates an impact due to the noise on the waveforms that affects the energy estimation of both methods in a similar way. The projections are shown for both axes. On the right, the distribution of residuals between both reconstruction methods is shown.}
\label{fig:FullE_Residual_train_CNN_Std}
\end{figure}

MC-based checks are performed for the exposure of the \mbox{EXO-200} detector with gamma rays emitted from a radioactive $^{228}$Th source at position S5. The correlation between the reconstructed energy from the neural network and the true MC energy is shown in Figure~\ref{fig:FullE_Prediction_train_CNN}. 
The energy reconstruction by the DNN works as expected. However, there is a small number of outlier events that are also present in the standard reconstruction (cf.\ Figure~\ref{fig:FullE_Prediction_train_Std}). The residuals to the true energy of both methods do not show any energy-dependent features other than broadening. The correlation between the residuals of both reconstruction methods with respect to the true MC energy is shown in Figure~\ref{fig:FullE_Residual_train_CNN_Std}. The positive correlation of these residuals indicates an impact due to the noise on the waveforms that affects the energy estimation of both methods in a similar way.

\begin{figure}[htp] 
    \centering
    \subfloat[$^{228}$Th spectra, all events]{\includegraphics[width=0.48\textwidth]{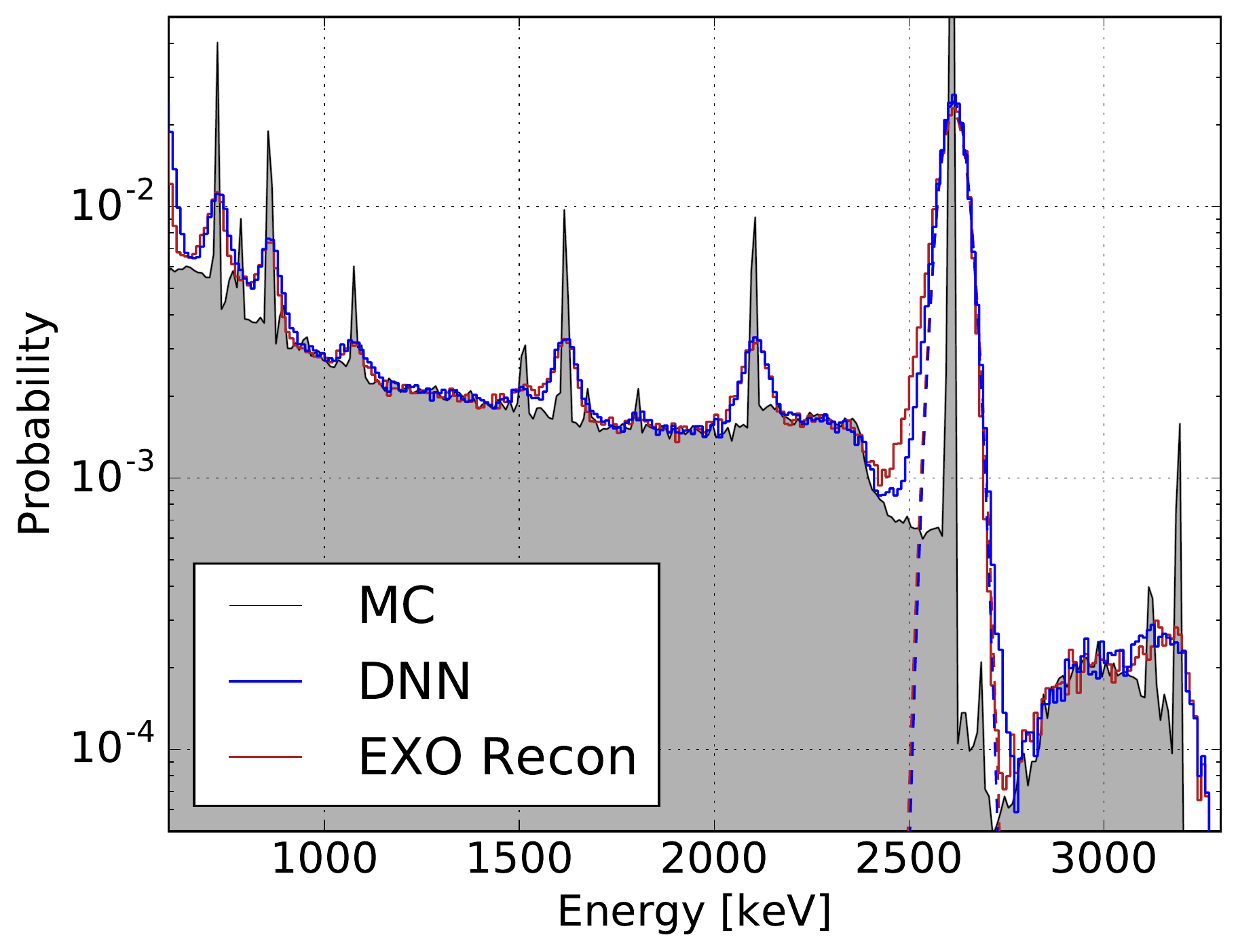}\label{fig:FullE_Spectrum_SSMS_train}}
    \qquad
    \subfloat[$^{228}$Th spectra, SS-only events]{\includegraphics[width=0.48\textwidth]{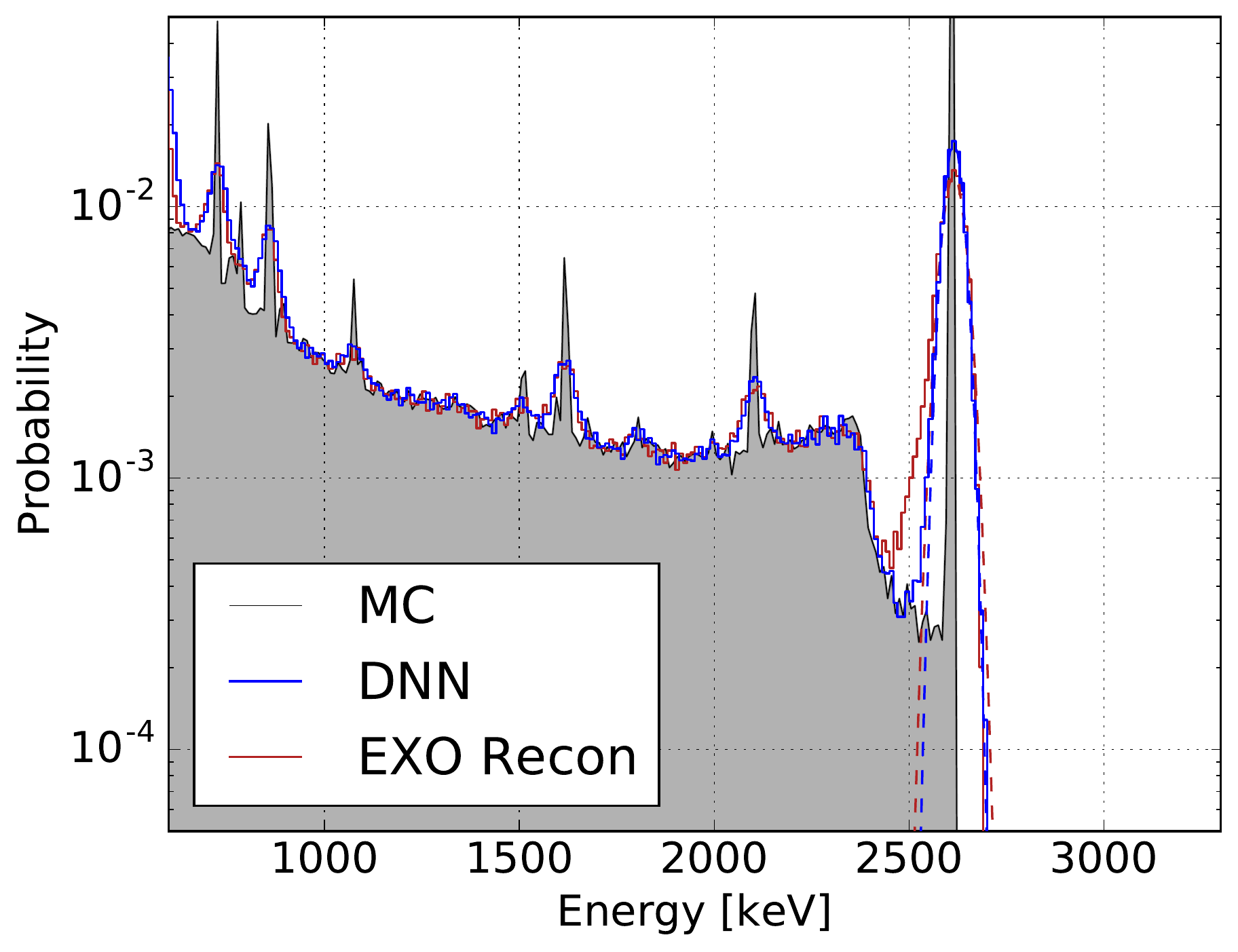}\label{fig:FullE_Spectrum_SS_train}}
    \quad
    \subfloat[$^{228}$Th spectra, MS-only events]{\includegraphics[width=0.48\textwidth]{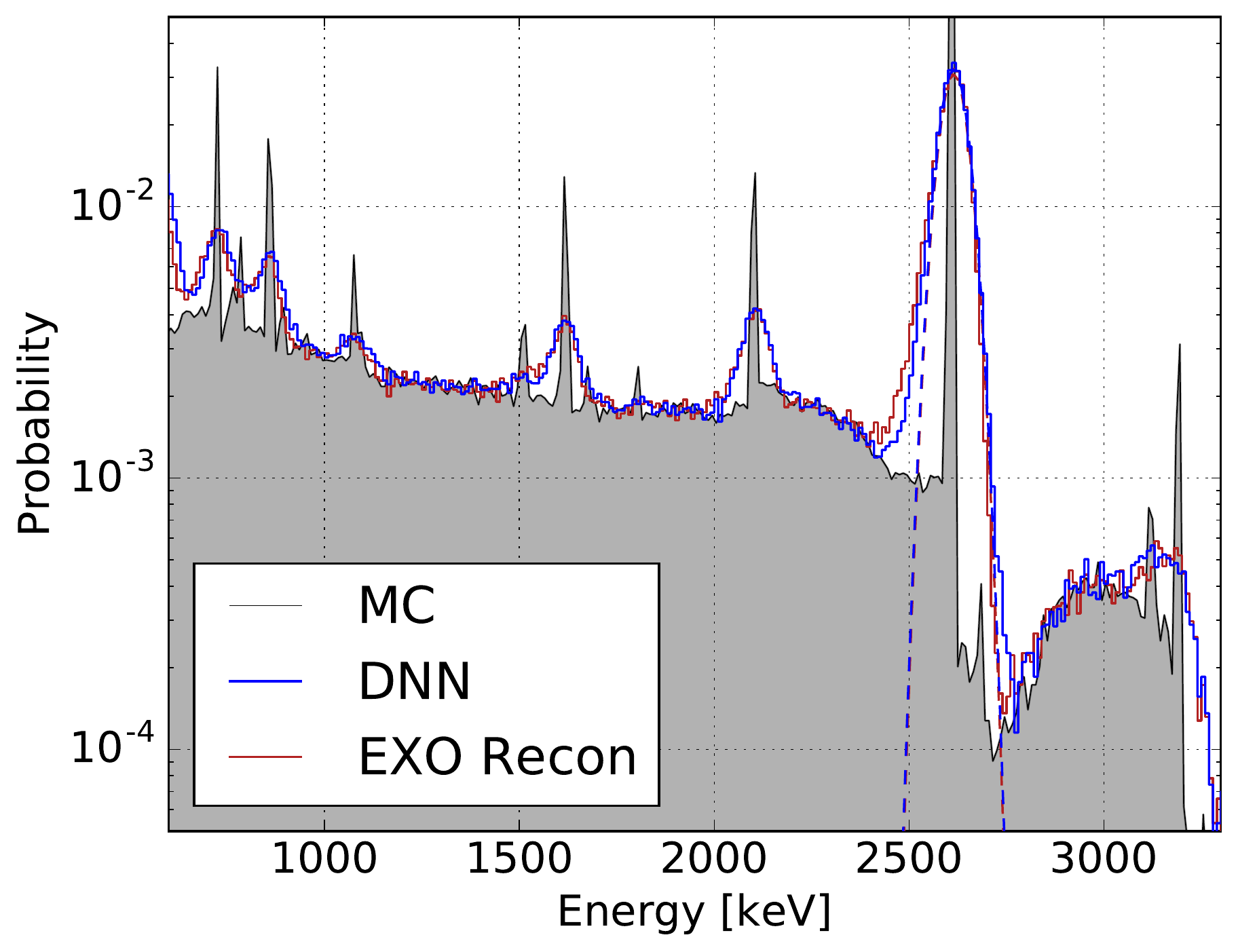}\label{fig:FullE_Spectrum_MS_train}}
    \caption{Energy spectra reconstructed from MC events from a $^{228}$Th calibration source with both the conventional reconstruction (red) and the neural network (blue) are shown in~\textbf{(a)}. The MC spectrum is shown in gray. \textbf{(b)} shows the spectra for SS-only events and~\textbf{(c)} shows spectra for MS-only events. A major contribution to the energy resolution is the electronics noise that is added to the MC signal. For SS-only events, the neural network performs noticeably better in the region between the Compton shoulder and the full absorption peak.}
    \label{fig:FullE_Spectrum_train} 
\end{figure}

Comparing the reconstructed $^{228}$Th energy spectra to the MC input spectrum, the spectral components, e.g.\ the full absorption peak at \SI{2614.5}{\kilo\electronvolt} from the decay of $^{208}$Tl and the single and double escape peaks, are reproduced by both reconstruction methods. Both methods yield a similar energy resolution at the full absorption peak at \SI{2.6}{\mega\electronvolt} (\SI{1.29+-0.03}{\percent} for EXO-recon and \SI{1.22+-0.02}{\percent} for DNN) where a major contribution is electronics noise on the waveforms. The energy resolution of SS-only events at that peak is \SI{1.15+-0.03}{\percent} (EXO-recon) and \SI{0.94+-0.01}{\percent} (DNN) whereas for MS-only events it is \SI{1.36+-0.04}{\percent} (EXO-recon) and \SI{1.38+-0.02}{\percent} (DNN) at that peak. The energy axis is calibrated in a simplified way compared to the conventional EXO analysis. Here, we apply a calibration factor for both SS and MS events separately, matching the $^{208}$Tl full absorption peak towards the nominal value of \SI{2614.5}{\kilo\electronvolt}. The reconstructed charge energy spectra for both the neural network and the conventional reconstruction are shown in Figure~\ref{fig:FullE_Spectrum_train} as well as the spectra for SS-only and MS-only events. The gray shaded spectra show the MC true energies. One noticeable difference between the neural network and the conventional reconstruction is the improved agreement in the region between the Compton shoulder and the full absorption peak. Two main processes contribute to this. First, the neural network is able to reconstruct smaller energy deposits that are below the reconstruction threshold of the conventional method. Second, collection signals superimposed by induction signals are partially flagged as pure induction signal by the conventional reconstruction whereas the neural network outperforms this by an improved capability to disentangle mixed signals.

\begin{figure}[h]
\centering 
\includegraphics[width=0.6\textwidth]{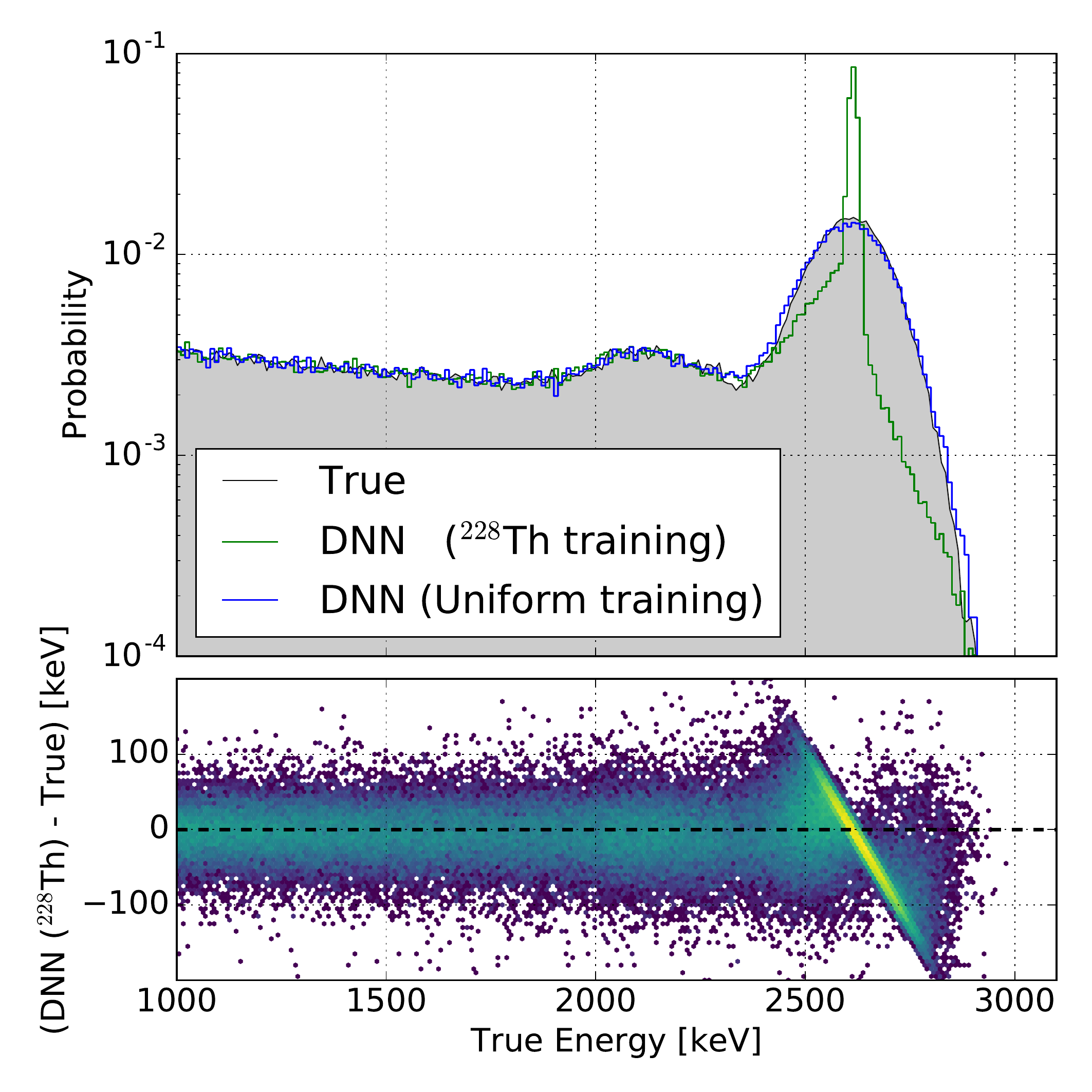}
\caption{The MC spectrum (gray) reconstructed by two different neural networks (top). The MC events are generated from a $^{228}$Th calibration source whose energy spectrum is broadened with an arbitrary, artificial resolution of \SI{3.5}{\percent}. One network was trained with a $^{228}$Th MC source (green) and the other network (blue) was trained on a uniform energy spectrum. The energy residuals of the network trained on the $^{228}$Th MC spectrum (shown in Figure~\ref{fig:FullE_Spectrum_SSMS_train} in gray) to the true MC energy is shown in the lower plot, where the color denotes the intensity on a logarithmic scale. The residuals indicate an anti-correlation that leads to an artificial improvement in resolution at the peak of the $^{208}$Tl decay at \SI{2.6}{\mega\electronvolt}.}
\label{fig:FullE_Residual_broad}
\end{figure}

Initially, in a manner analogous to testing, training was performed on MC events by gamma rays emitted from a radioactive $^{228}$Th calibration source. Both training and validation results were promising. Testing the network's performance on MC events from a $^{228}$Th calibration source whose energy spectrum was broadened with an arbitrary, artificial resolution of \SI{3.5}{\percent} (guided by the charge-only energy resolution from \mbox{EXO-200}) revealed undesired issues. The reconstructed energy of events from the full absorption peak of the $^{208}$Tl decay at around \SI{2.6}{\mega\electronvolt} is shuffled toward the nominal value of the peak, shown in Figure~\ref{fig:FullE_Residual_broad} (green). Intuitively, this can be explained by the network's smaller loss contribution during training, when events from around the peak are reconstructed. Compared to events around this peak, the abundance of $^{208}$Tl peak events is larger by orders of magnitude (c.f.\ Figure~\ref{fig:FullE_Spectrum_train}). Thus, on average, the contribution is smaller when reconstructing events toward the peak than away from it. This mimics a seemingly improved resolution at the $^{208}$Tl peak. Hence, from this point onwards, a training spectrum uniform in energy was utilized to improve the network's capability of reconstructing events from different energy distributions. Using this training approach, the broadened spectrum is reconstructed correctly without introducing systematic effects as shown in Figure~\ref{fig:FullE_Residual_broad} (blue).

\subsection{Testing on calibration data}
\begin{figure}[htp!]
\centering 
\includegraphics[width=0.6\textwidth]{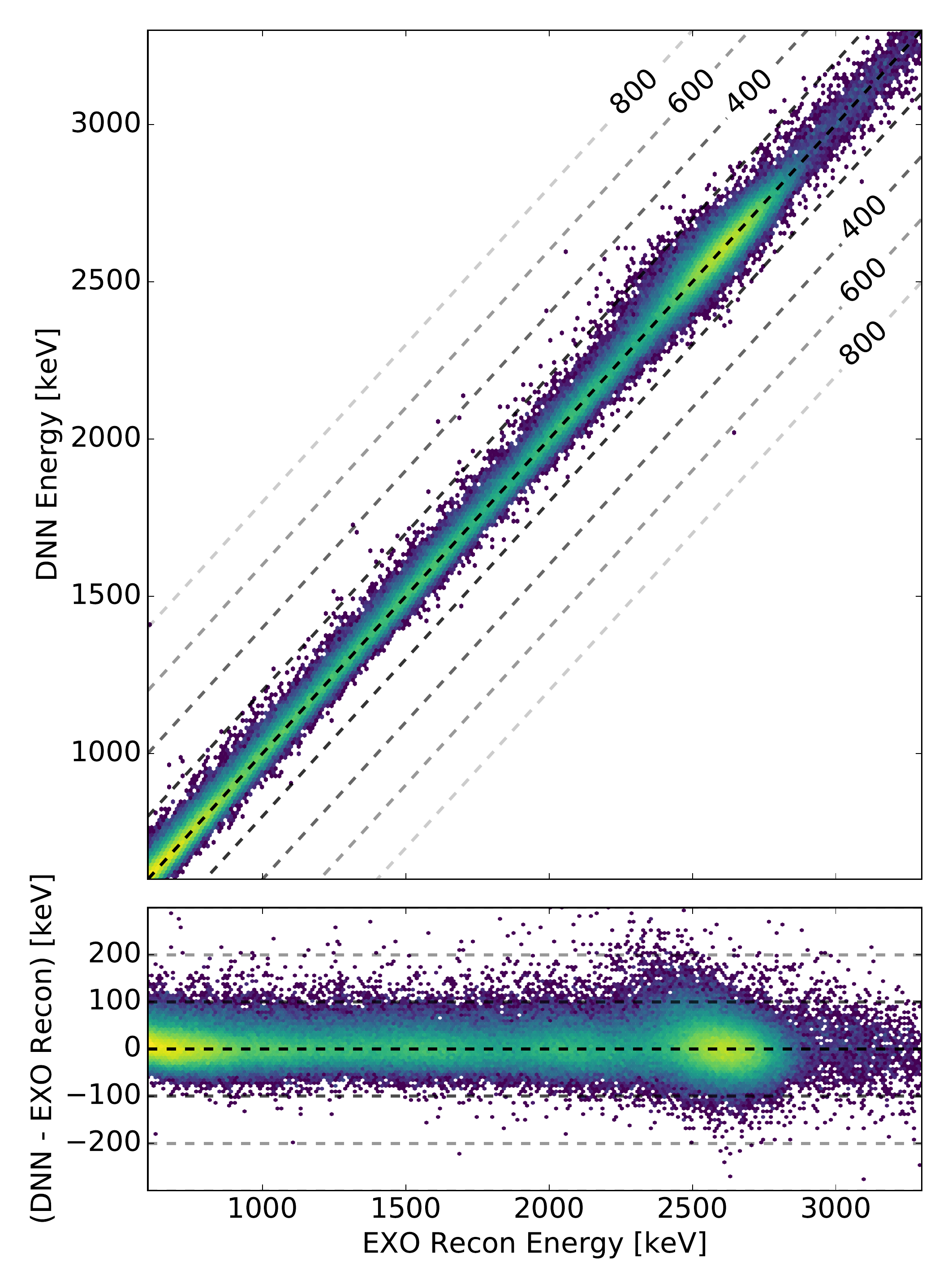}
\caption{Energies of experimental events from a $^{228}$Th calibration source at position S5 reconstructed by the neural network as function of the energy reconstructed by the conventional analysis. The corresponding residuals are shown in the lower panel. The color denotes the intensity on a logarithmic scale. Dashed lines indicate a certain residual as given in the figure. The residual of both methods does not show systematic energy dependent features.}
\label{fig:FullE_Prediction_Th_Data_S5}
\end{figure}

Performance checks with real data events are done with $^{228}$Th events of a calibration source at position S5 from \mbox{EXO-200} data taken in Phase~I. The event selection is analogous to the MC event selection introduced in Section~\ref{sec:FullE_Train}. Additionally, events with multiple scintillation clusters (pile-up of light signals) are discarded to ensure that the charge signal is in the central \SI{1024}{\micro\second} window that is used in this study. The correlation between the reconstructed energy from the neural network and the conventional reconstruction is shown in Figure~\ref{fig:FullE_Prediction_Th_Data_S5}. The difference in reconstructed energy between both methods does not show any energy dependent features other than broadening.

The time variation of the LXe purity affects the electron lifetime and thus the detector's response to a given initial energy deposit. During Phase~I of \mbox{EXO-200}, a mean electron lifetime of \SI{4.5}{\milli\second} was achieved~\cite{PhysRevC.89.015502}. This value is used for generating MC events. However, the electron lifetime varies between different runs used for testing the network on real data. For consistency, the results discussed here are produced from runs with an electron lifetime of about \SI{4.5}{\milli\second}. However, we do not see any loss in performance of the network on events with a very different electron lifetime.

\begin{figure}[h]
\centering 
\includegraphics[width=0.6\textwidth]{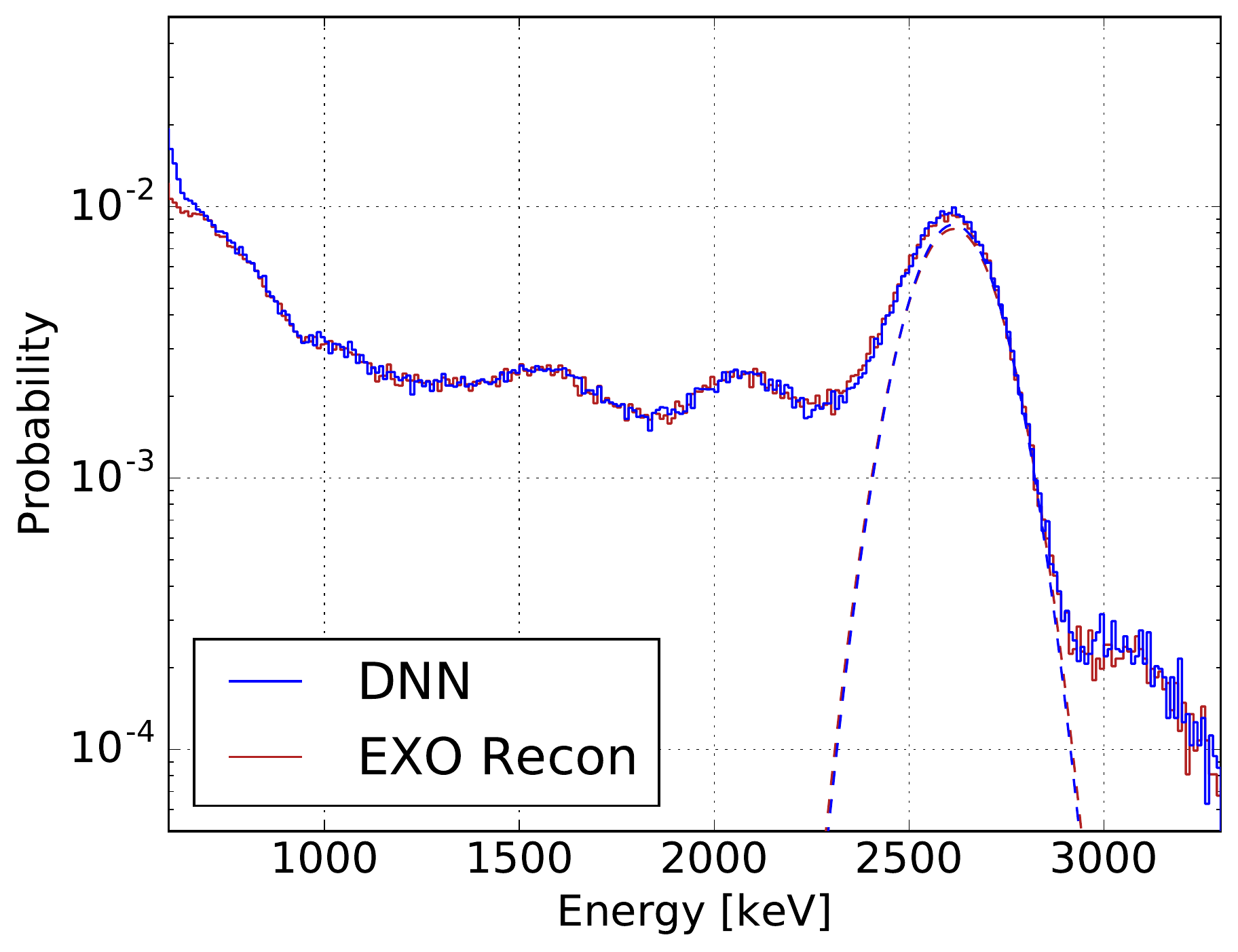}
\caption{Charge energy spectra reconstructed from data events from a $^{228}$Th calibration source at position S5 with both the conventional analysis (red) and the neural network (blue). The energy axis is calibrated as discussed in Section~\ref{sec:FullE_Train}. The resolution is dominated by the recombination fluctuation between the signatures in the light and the charge channel.}
\label{fig:FullE_Spectrum_Th_Data_S5}
\end{figure}

The neural network yields competitive results compared to the conventional reconstruction, seen in Figure~\ref{fig:FullE_Spectrum_Th_Data_S5}. The energy resolution at the \SI{2.6}{\mega\electronvolt} full absorption peak of the $^{208}$Tl decay is similar for both methods (\SI{3.93+-0.04}{\percent} for EXO-recon and \SI{3.84+-0.03}{\percent} for DNN). The energy resolution of SS-only events at that peak is \SI{3.53+-0.07}{\percent} (EXO-recon) and \SI{3.44+-0.06}{\percent} (DNN) whereas for MS-only events it is \SI{4.08+-0.04}{\percent} (EXO-recon) and \SI{4.02+-0.03}{\percent} (DNN) at that peak. The energy axis is calibrated following the procedure introduced in Section~\ref{sec:FullE_Train} for the MC case.

\begin{figure}[htp!] 
    \centering
    \subfloat[$^{228}$Th source at S2]{\includegraphics[width=0.48\textwidth]{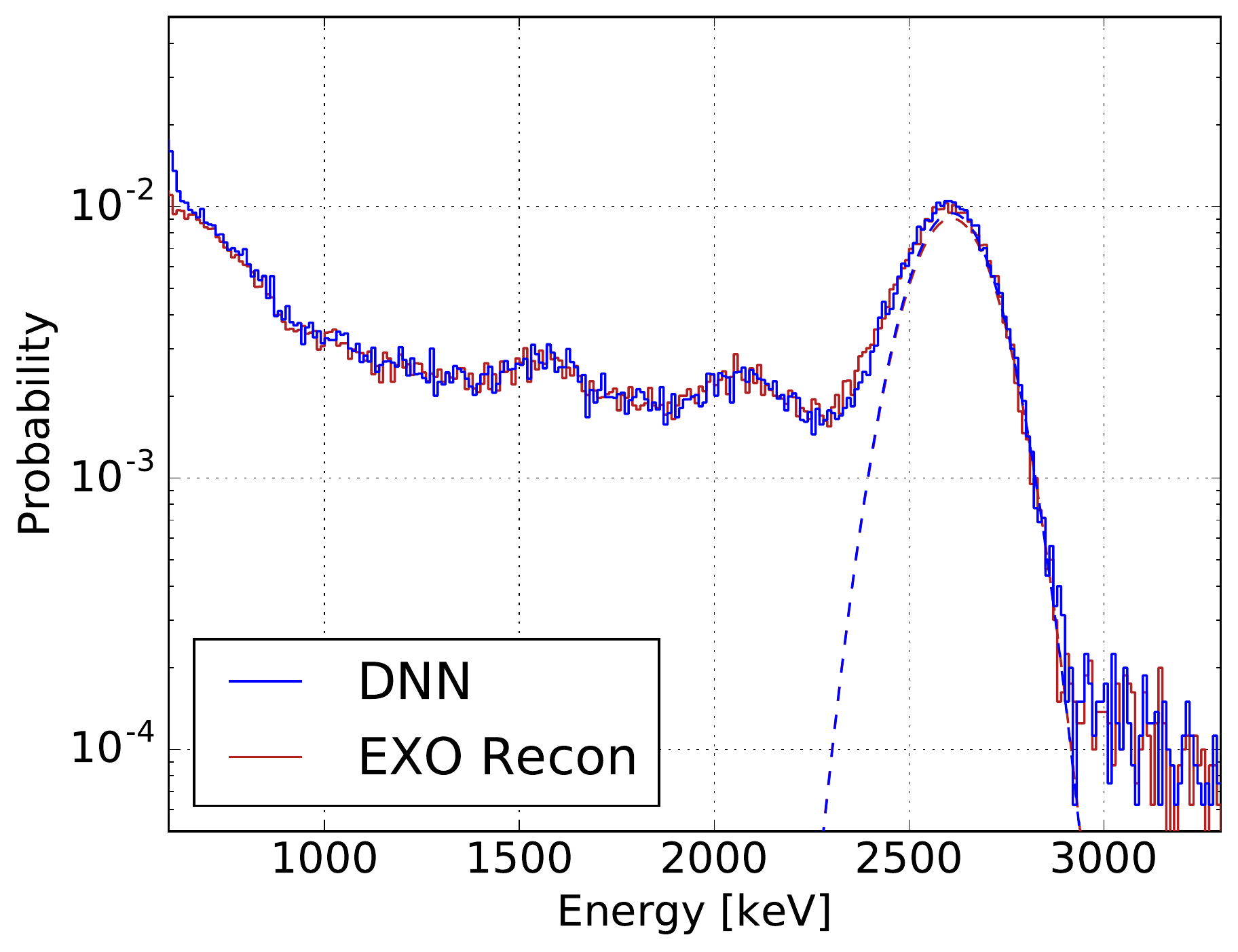}\label{fig:FullE_SpectrumRot_Th_Data_S2}}
    \quad
    \subfloat[$^{228}$Th source at S8]{\includegraphics[width=0.48\textwidth]{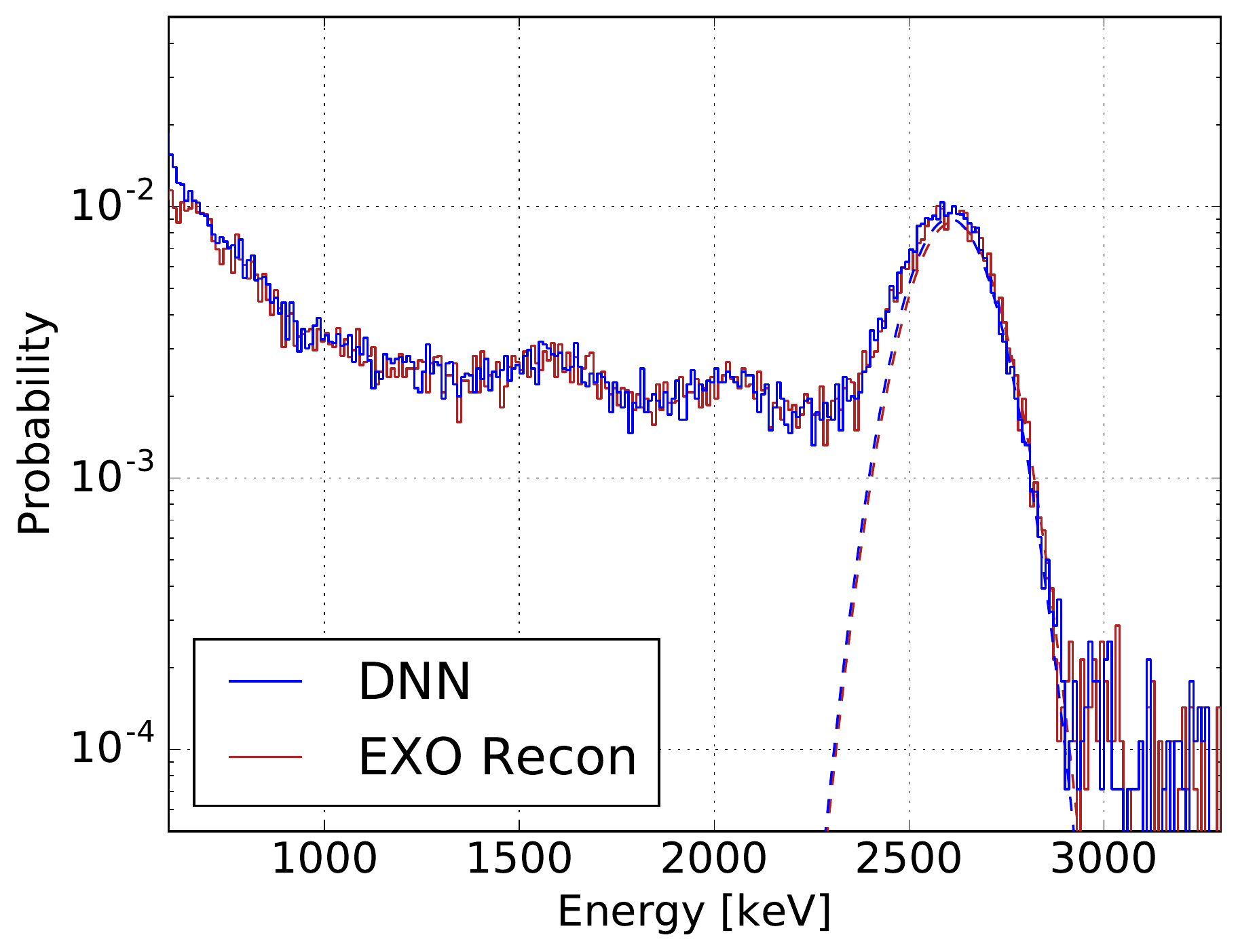}\label{fig:FullE_SpectrumRot_Th_Data_S8}}
    \caption{Energy spectra reconstructed from data events from a $^{228}$Th calibration source at calibration positions S2~\textbf{(a)} and S8~\textbf{(b)} with both the conventional analysis (red) and the neural network (blue). The resolution is dominated by recombination fluctuations between the light and the charge channel.}
    \label{fig:FullE_SpectrumRot_Th_Data_S2_S8} 
\end{figure}

The network's performance on events of the $^{228}$Th calibration source when located at other positions is also evaluated. Even though the training is performed on a MC spectrum uniform in energy, the events are not uniformly distributed over the detector volume but are concentrated near the cathode for training and for source position S5. Thus, the position distribution of calibration events from position S5 is similar to the one from the MC gamma ray source. To test the network's performance on highly different event distributions, events from $^{228}$Th runs at positions S2 and S8 are evaluated. These positions are centered on the drift axis behind either anode plane. The resulting energy spectra for the conventional reconstruction and the neural network are shown in Figure~\ref{fig:FullE_SpectrumRot_Th_Data_S2_S8}. Again, the energy resolution at the \SI{2.6}{\mega\electronvolt} peak from the $^{208}$Tl decay is similar for the two methods. We conclude that the position distribution does not have a significant impact on the network's performance in reconstructing energy.

\begin{figure}[t!] 
    \centering
    \subfloat[\label{fig:FullE_Anticorr_Th_Data_S5_CNN}]{\includegraphics[width=0.48\textwidth]{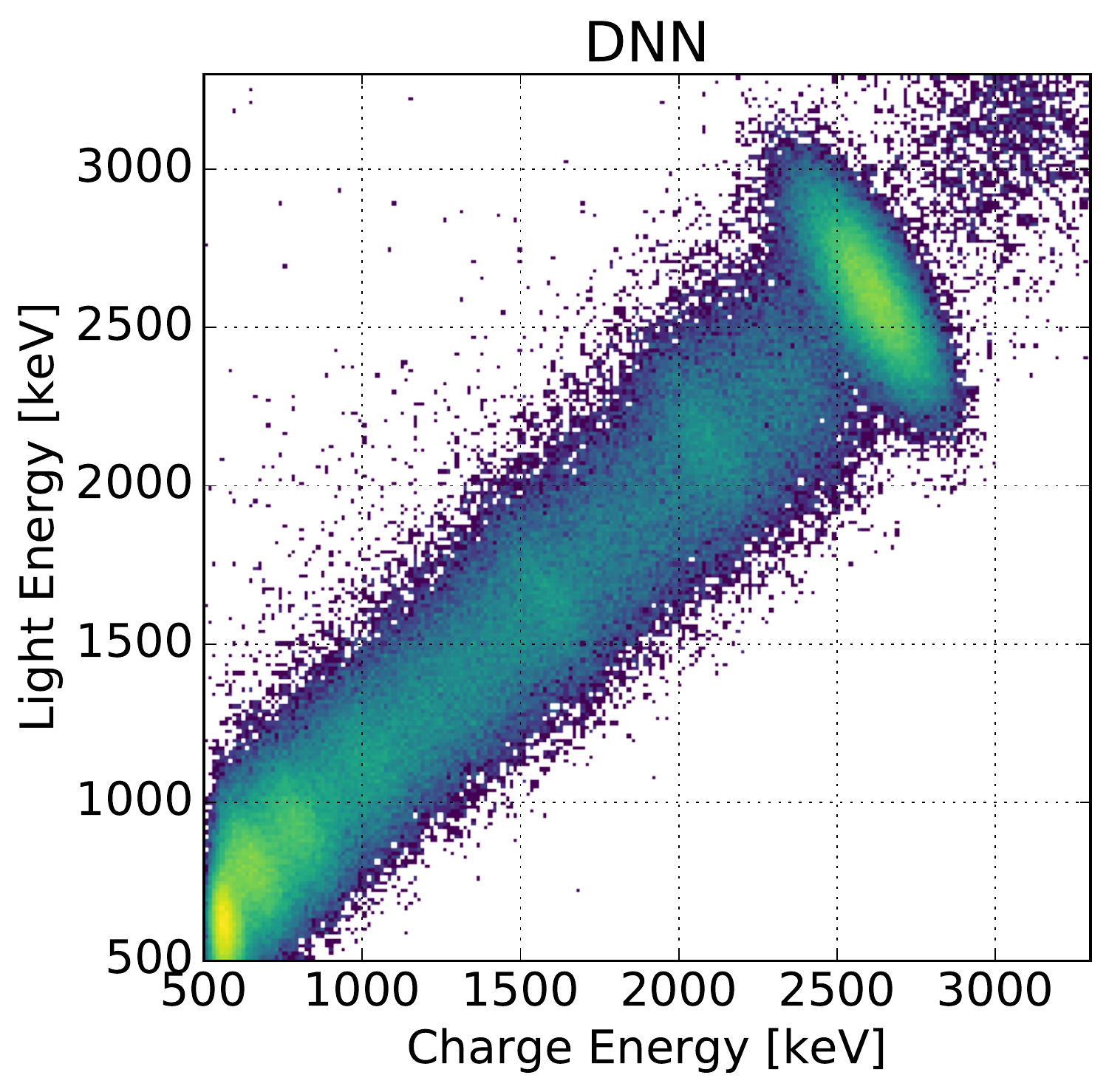}}
    \quad
    \subfloat[\label{fig:FullE_Anticorr_Th_Data_S5_Std}]{\includegraphics[width=0.48\textwidth]{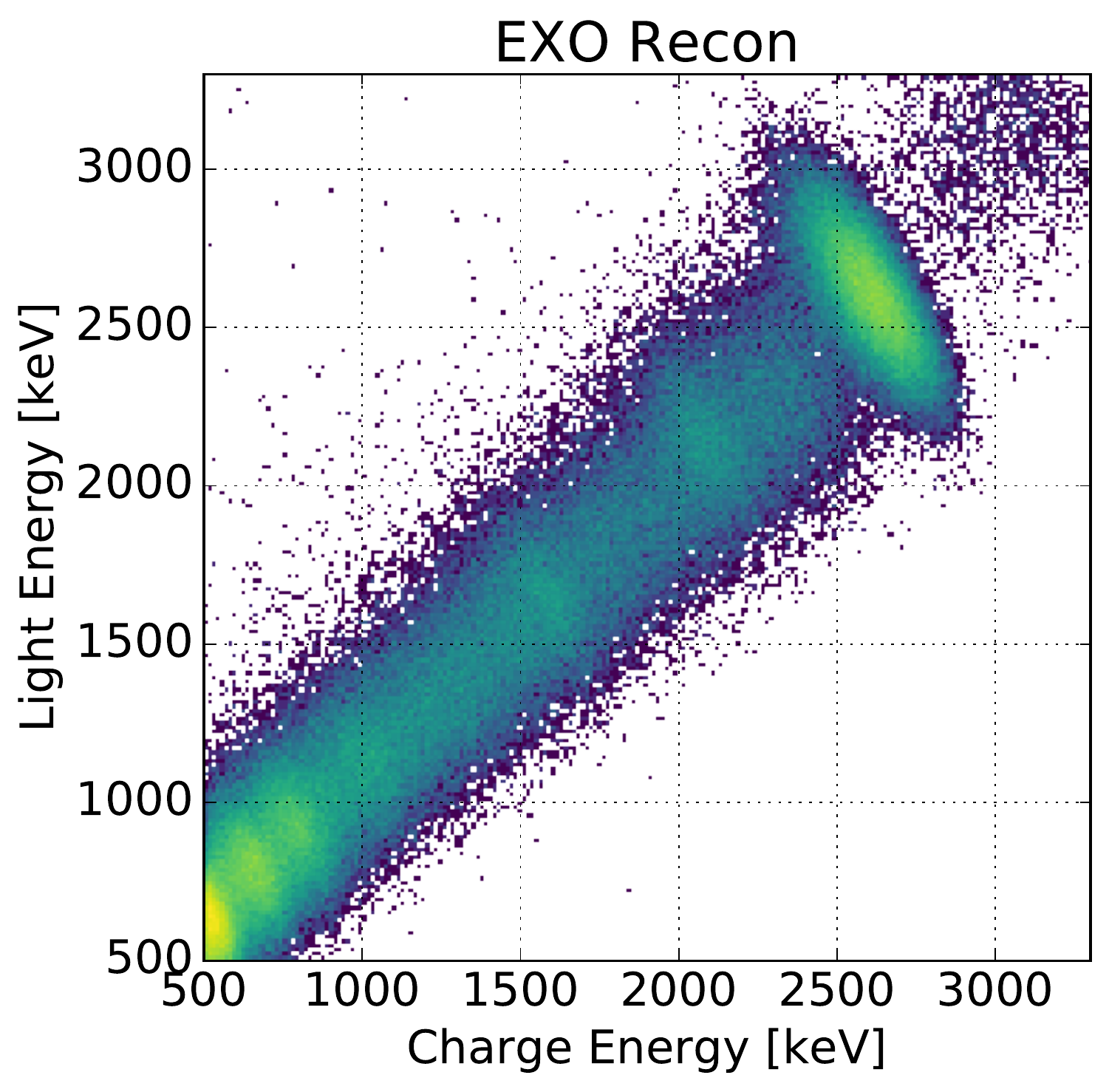}}
    \caption{Anti-correlation between ionization and scintillation from a $^{228}$Th calibration source. The ionization energy is reconstructed by the neural network (left) and by the conventional analysis (right) whereas the scintillation energy is given by the conventional analysis. The color denotes the intensity on a logarithmic scale. The prominent feature is the full absorption peak of $^{208}$Tl.}
    \label{fig:FullE_Anticorr_Th_Data_S5} 
\end{figure}

\begin{figure}[htp!]
	\centering
    \subfloat[$^{228}$Th spectra, all events]{\includegraphics[width=0.48\textwidth]{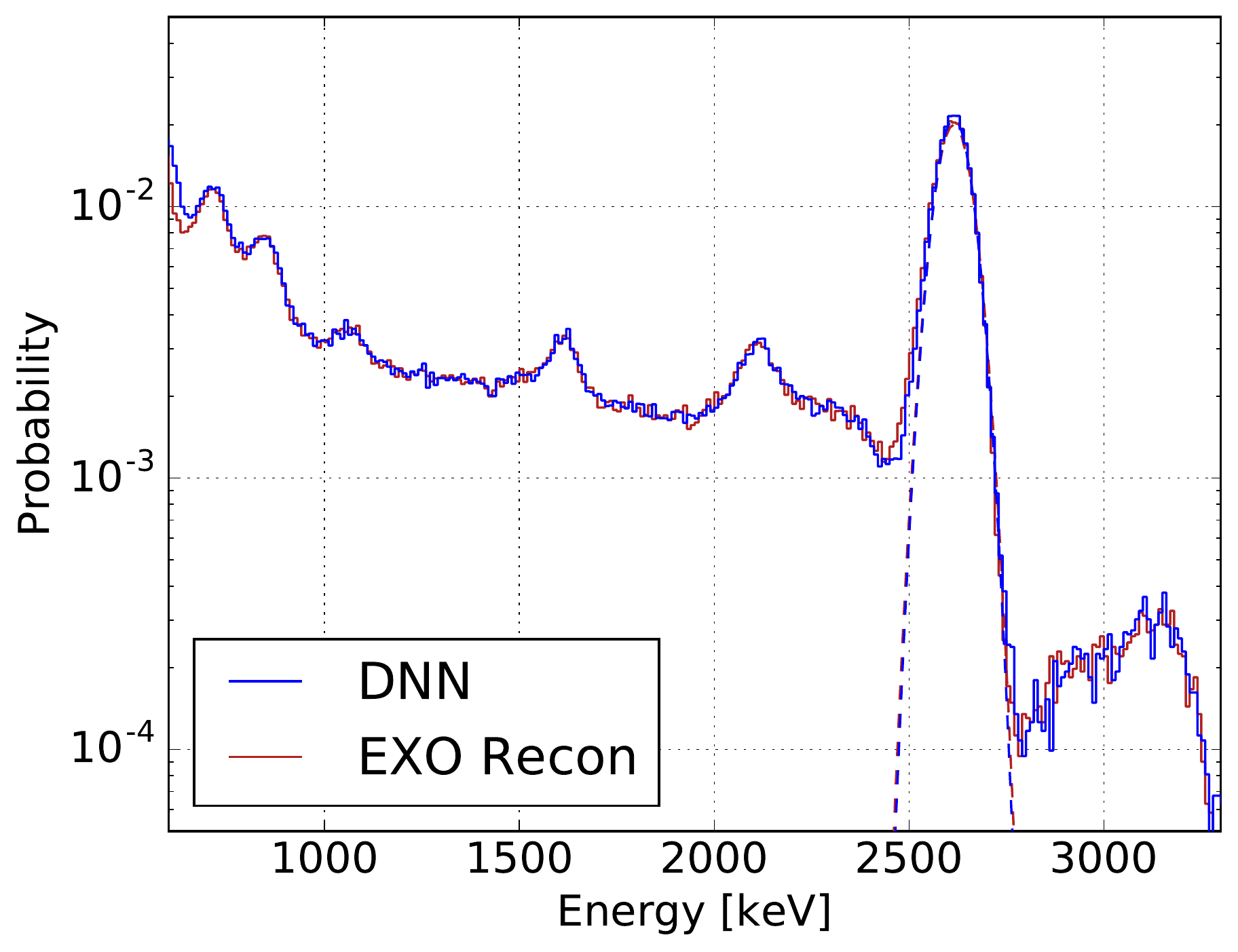}\label{fig:FullE_SpectrumRot_Th_Data_S5_SSMS}}
    \qquad
    \subfloat[$^{228}$Th spectra, SS-only events]{\includegraphics[width=0.48\textwidth]{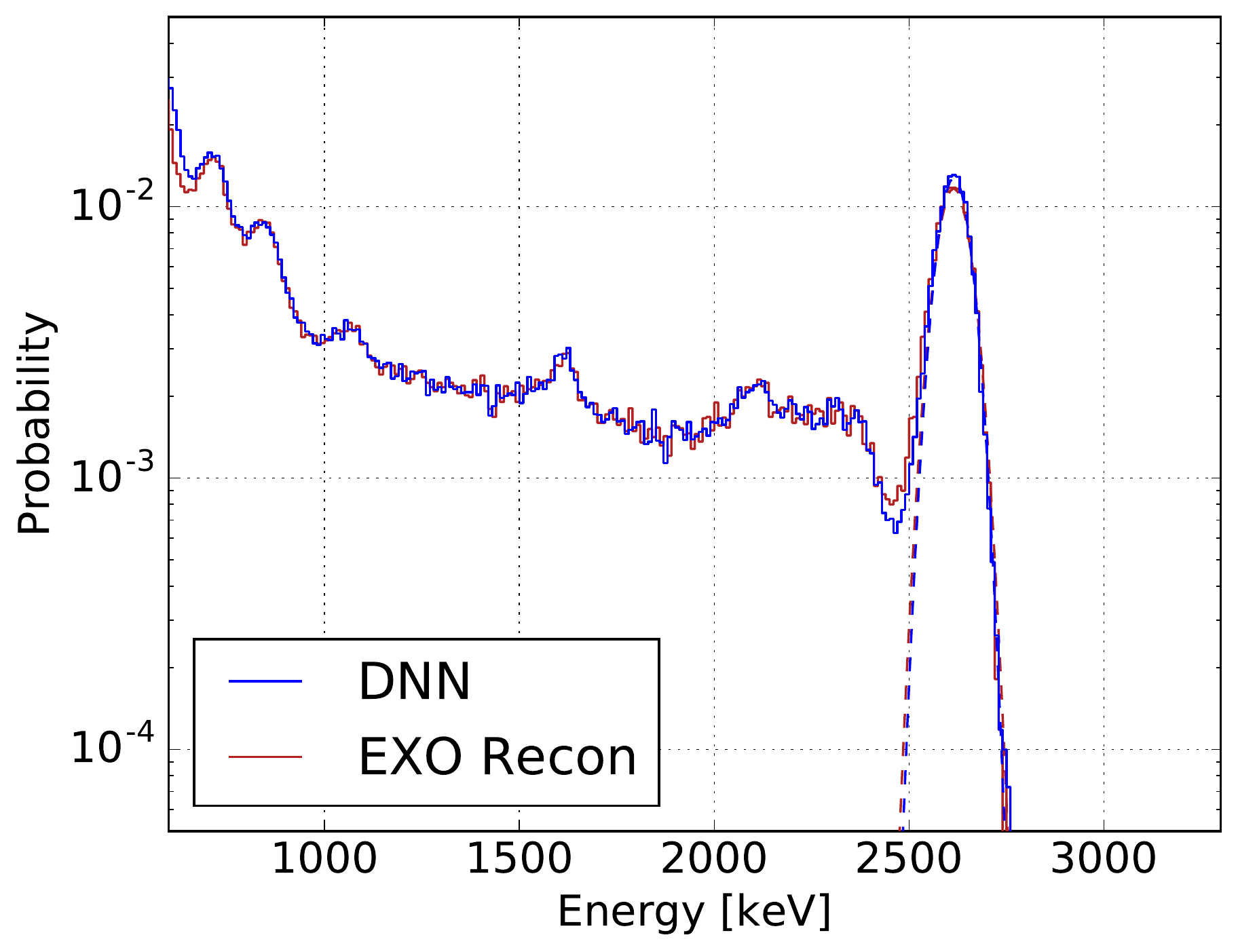}\label{fig:FullE_SpectrumRot_Th_Data_S5_SS}}
    \quad
    \subfloat[$^{228}$Th spectra, MS-only events]{\includegraphics[width=0.48\textwidth]{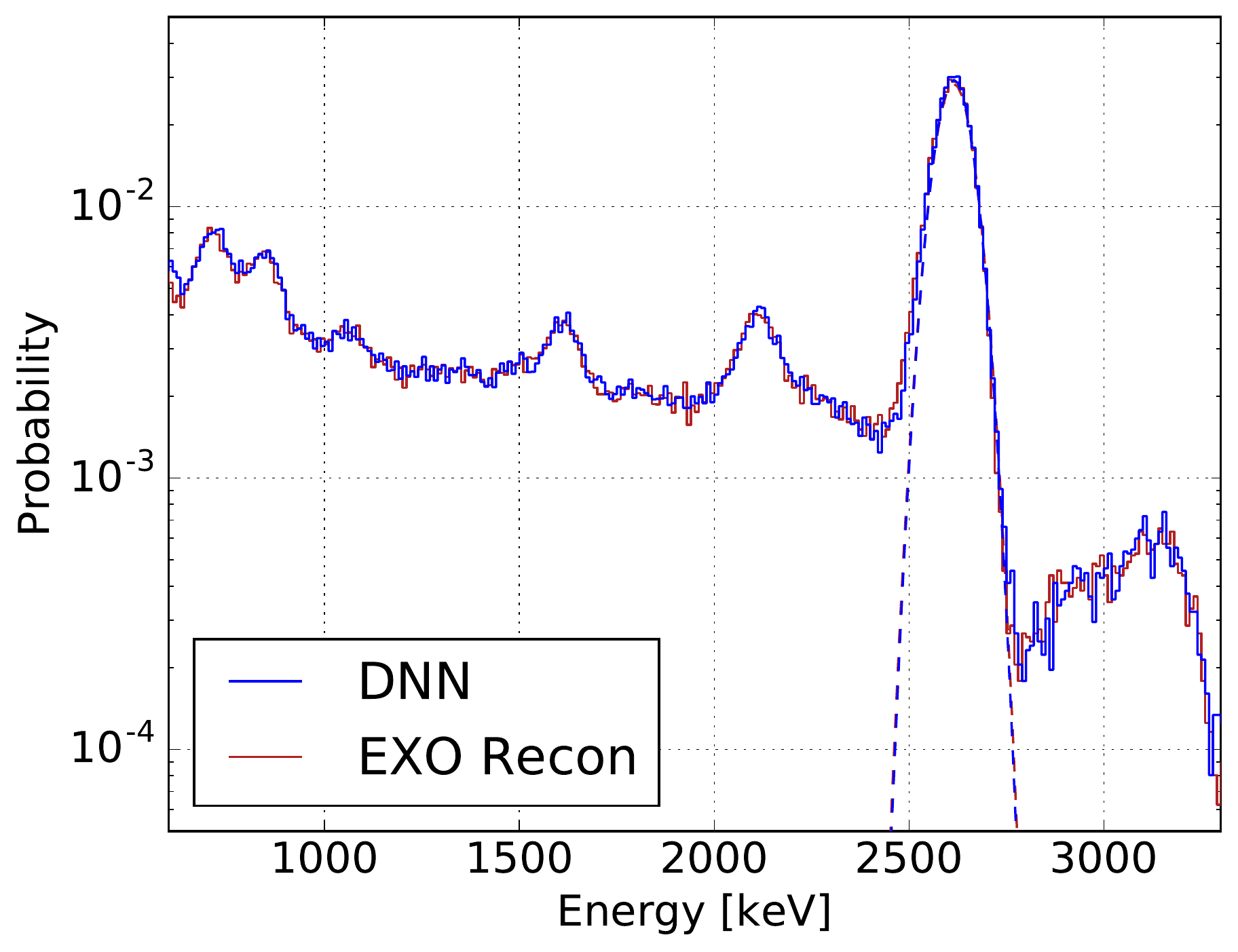}\label{fig:FullE_SpectrumRot_Th_Data_S5_MS}}
    \caption{Reconstructed energy spectra after combining the ionization and the scintillation channel with the optimal linear combination from data events of a $^{228}$Th calibration source at position S5 with both the conventional analysis (red) and the neural network (blue) are shown in~\textbf{(a)}. \textbf{(b)} shows the spectra for SS-only events and~\textbf{(c)} shows spectra for MS-only events. The energy axis is calibrated as discussed in Section~\ref{sec:FullE_Train}. The shapes of both spectra of the conventional analysis (red) and the one by the neural network (blue) agree and yield a similar resolution at the full absorption peak of the $^{208}$Tl decay at \SI{2.6}{\mega\electronvolt}. For SS-only events, the neural network performs noticeably better in the region between the Compton shoulder and the full absorption peak.}
\label{fig:FullE_SpectrumRot_Th_Data_S5}
\end{figure}

Due to the anti-correlation of ionization and scintillation in LXe, the energy resolution can be improved by combining both signatures. This is illustrated in Figure~\ref{fig:FullE_Anticorr_Th_Data_S5}. 
For the purpose of comparing the network's performance with the conventional EXO reconstruction, we utilize the scintillation signal assessed by the conventional EXO reconstruction for both methods. To improve the energy resolution, denoising is applied to the scintillation channel as introduced in~\cite{Davis:2016reu}. The optimal energy variable is found by determining the linear combination of charge and light signals that minimizes the energy resolution at the $^{208}$Tl peak. This linear combination is applied to all events regardless of their energy~\cite{PhysRevC.89.015502}. This is done for SS and MS events separately. The resulting energy spectra are shown in Figure~\ref{fig:FullE_SpectrumRot_Th_Data_S5} and yield comparable results for both reconstruction methods of the charge energy with an energy resolution at the $^{208}$Tl peak of \SI{1.70+-0.02}{\percent} (EXO-recon) and \SI{1.65+-0.02}{\percent} (DNN). The energy resolution of SS-only events at that peak is \SI{1.61+-0.03}{\percent} (EXO-recon) and \SI{1.50+-0.02}{\percent} (DNN) whereas for MS-only events it is \SI{1.75+-0.02}{\percent} (EXO-recon) and \SI{1.71+-0.02}{\percent} (DNN) at that peak. This indicates a physically valid reconstruction by the neural network.

\begin{figure}[htp] 
    \centering
    \subfloat[$^{226}$Ra spectra]{\includegraphics[width=0.48\textwidth]{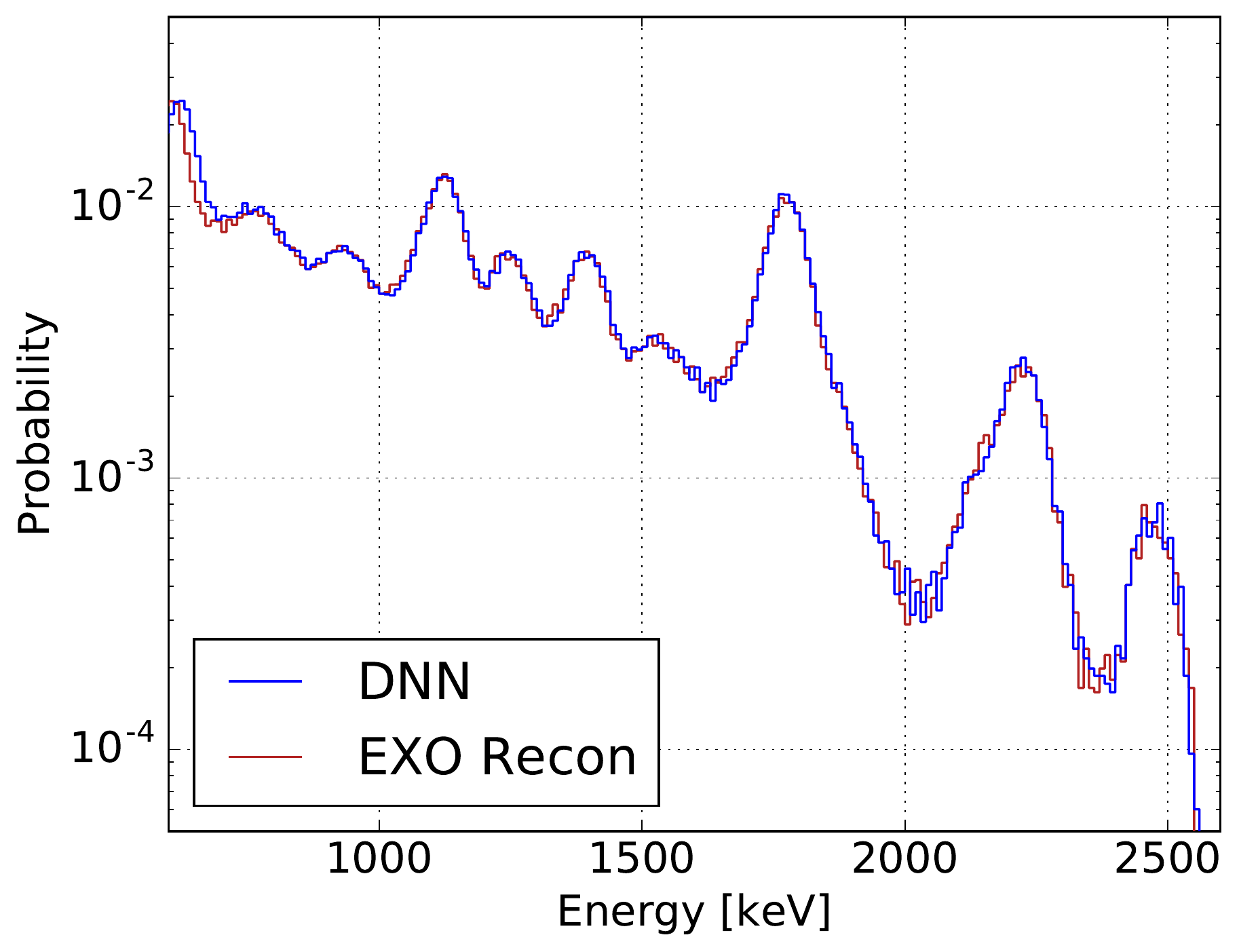}\label{fig:FullE_SpectrumRot_Ra_Data_S5}}
    \quad
    \subfloat[$^{60}$Co spectra]{\includegraphics[width=0.48\textwidth]{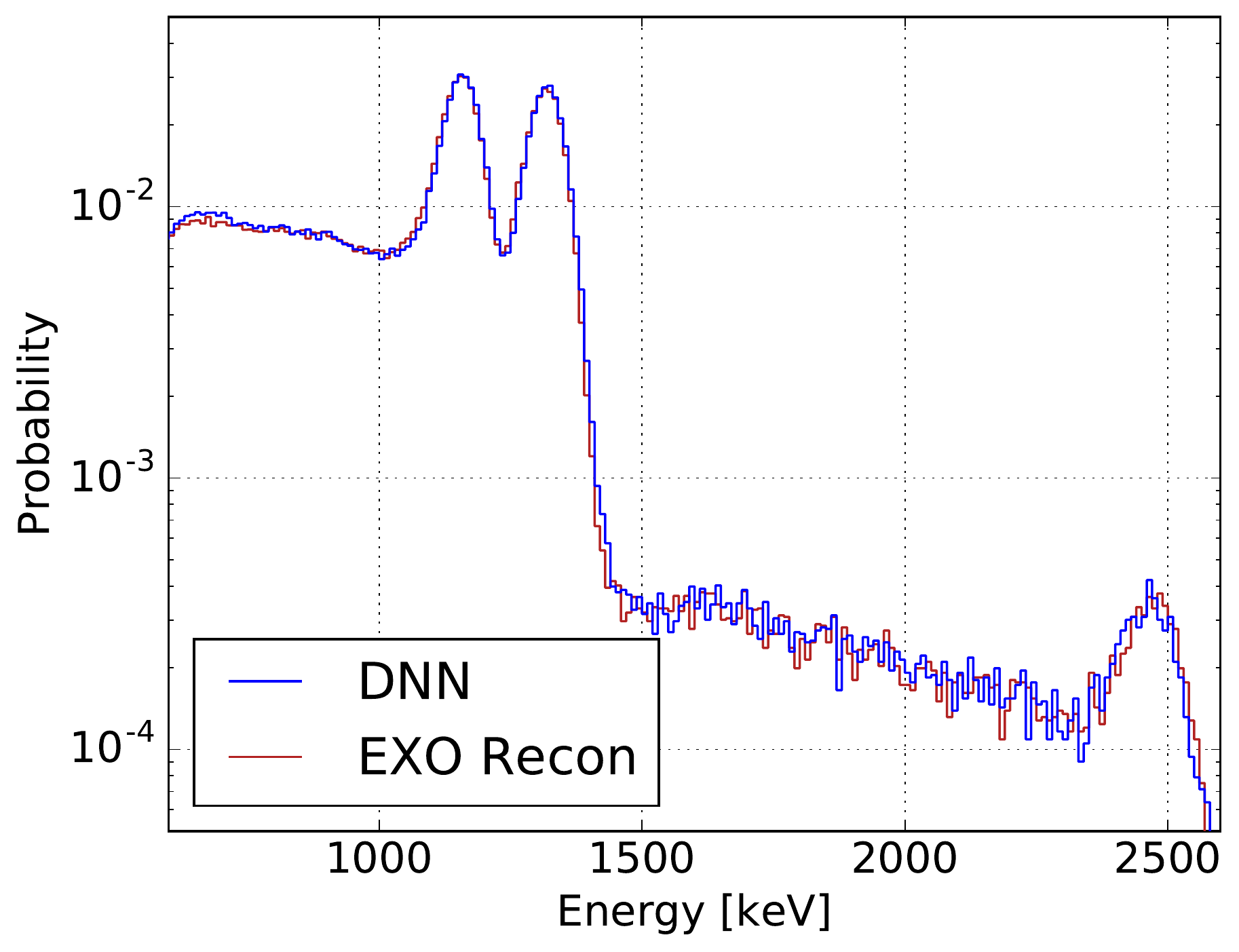}\label{fig:FullE_SpectrumRot_Co_Data_S5}}
    \caption{Reconstructed energy spectra from combining the ionization and the scintillation channel from data events from a $^{226}$Ra~\textbf{(a)} and a $^{60}$Co~\textbf{(b)} calibration source at source position S5. The shapes of the spectra of the conventional analysis (red) and the one by the neural network (blue) agree.}
    \label{fig:FullE_SpectrumRot_Ra_Co_Data_S5} 
\end{figure}

In order to probe different energies, the same procedure of combining the scintillation and ionization signatures was applied to events from a $^{226}$Ra and a $^{60}$Co calibration source at position S5. The corresponding spectra are shown in Figure~\ref{fig:FullE_SpectrumRot_Ra_Co_Data_S5}. The scintillation and ionization channels are combined with the same optimum linear combination as for $^{228}$Th. The neural network yields competitive results for all tested calibration sources with the shapes of the spectra being in good agreement with the conventional reconstruction.

\section{Data driven scintillation light reconstruction}
The previously discussed studies attempt to reconstruct the energy deposited by an event occurring in the detector through the amount of free electrons. To accomplish this task, deep neural networks are trained using MC simulations of such events. The accuracy and choice of simulated events will impact the performance of the reconstruction. The fact that \mbox{EXO-200} registers interactions of ionizing radiation not only through the collection of electrons but also by detecting the scintillation light from re-combinations and excitation, allows for novel approaches of training deep neural networks that do not rely on MC simulations. Here we describe one such approach to reconstruct the position of an event from the distribution of scintillation light signals among the \num{74} APD channels using a deep neural network, which is trained on the information extracted from the collected ionization. The ability to reconstruct the event position from scintillation light is valuable where insufficient charge collection leads to missing information. This may happen when the position of an event is close to the Teflon reflector, and part or all of the ionization charge is lost and therefore below the detection threshold. This is also often the case for $\alpha$-decay events which have a large light yield but small charge yield. In the conventional \mbox{EXO-200} analysis such events are excluded since for them no full 3D position can be reconstructed and they may have the wrong charge/light ratio.

In total, the \mbox{EXO-200} detector contains \num{468} APDs, which are read out in groups of \num{7}, resulting in \num{37} channels in each half of the TPC. They are arranged to cover a maximal area at the end caps of the cylindrical detector and are positioned behind the U- and V-Wires (for detailed description of the detector design see Ref.~\cite{Auger:2012gs} and Section~\ref{sec:detector}). The conventional \mbox{EXO-200} analysis extracts the X-Y position information of an event by combining the signal information from charge collection on the U-Wires and induction on the V-Wires. Together with the light signal and the known drift velocity, this enables a three dimensional position reconstruction with an accuracy of approximately $\sigma_{3D}$~=~\SI{3}{\milli\meter}. The limitation of the position resolution is dominated by the \SI{9}{\milli\meter} wire pitch between charge readout channels. When the electron cloud is large enough to be collected on more than one neighboring wire, an amplitude weighted position is used. Since we are using this position information with finite resolution to train the network, it also sets the best performance of the model. However, due to larger noise on the APD channels, their coarser segmentation compared to wire channels, and other physical limitations due to light absorption and reflection on different materials, the position resolution extracted from the scintillation light is expected to be worse. This further validates the use of charge information as truth labels and avoids the requirement to simulate light propagation and detection in the detector.
\subsection{Training data preparation}
\begin{figure}[htp]
\includegraphics[width=\textwidth]{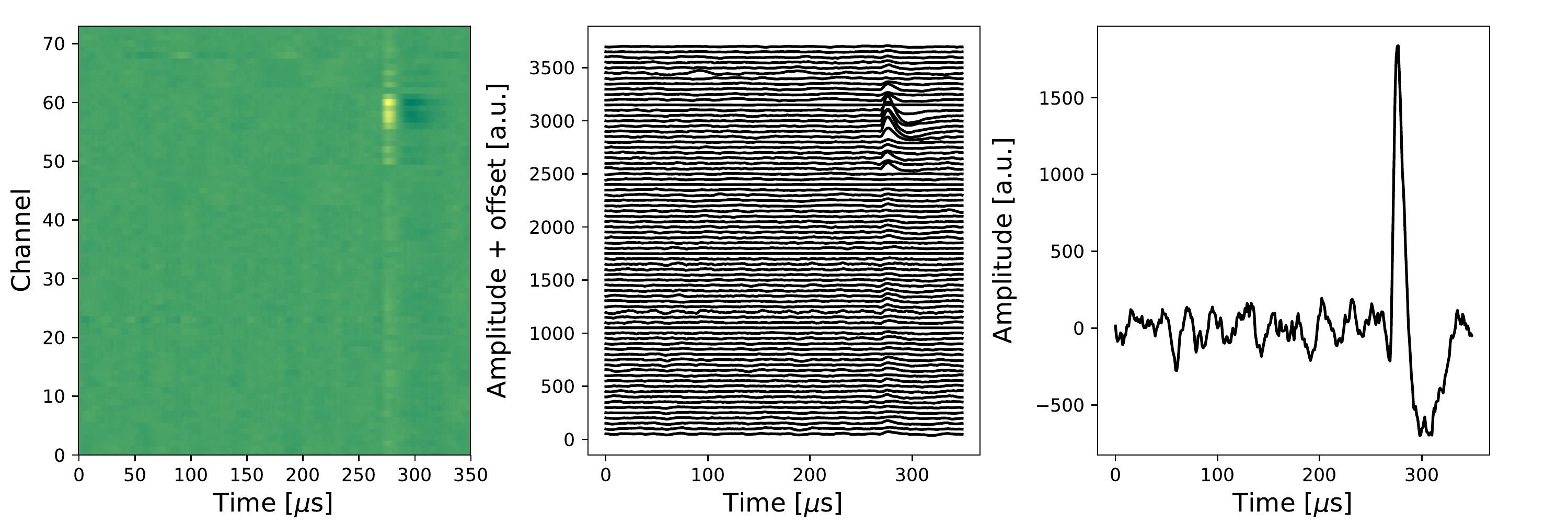}
\caption{Example of the input image to the convolutional neural network (left panel). The vertical axis represents the APD channel and the horizontal axis represents the time in \si{\micro\second}. The middle panel shows the waveforms of each APD channel while the right panel is the sum waveform over all channels.}
\label{fig:ScintRec_Image}
\end{figure}

The input data for the deep neural network consists of the raw waveforms of all \num{74} APD channels. The only pre-processing done is the subtraction of the baseline for each waveform determined from the first \num{680} samples, being the first third of the full waveform, on an event by event basis. Furthermore, we crop the waveforms to \num{350} samples which guarantees that, independent of whether the event was triggered on U/V-Wires or APDs, the scintillation signal is fully contained in the frame. All waveform signals are stacked vertically in the order of their assigned channel number to form an image of $\num{350}\times\num{74}$ pixels. Figure~\ref{fig:ScintRec_Image} (left panel) shows an example image where the event position is close to the anode plane of the first TPC half defined as positive Z. It illustrates how the position information is encoded in the image: the APDs in proximity to the event location collect the major part of the light and therefore have the largest signal amplitude. APDs far away from the event location have much smaller signal amplitudes.

\begin{figure}[htp]
\includegraphics[width=\textwidth]{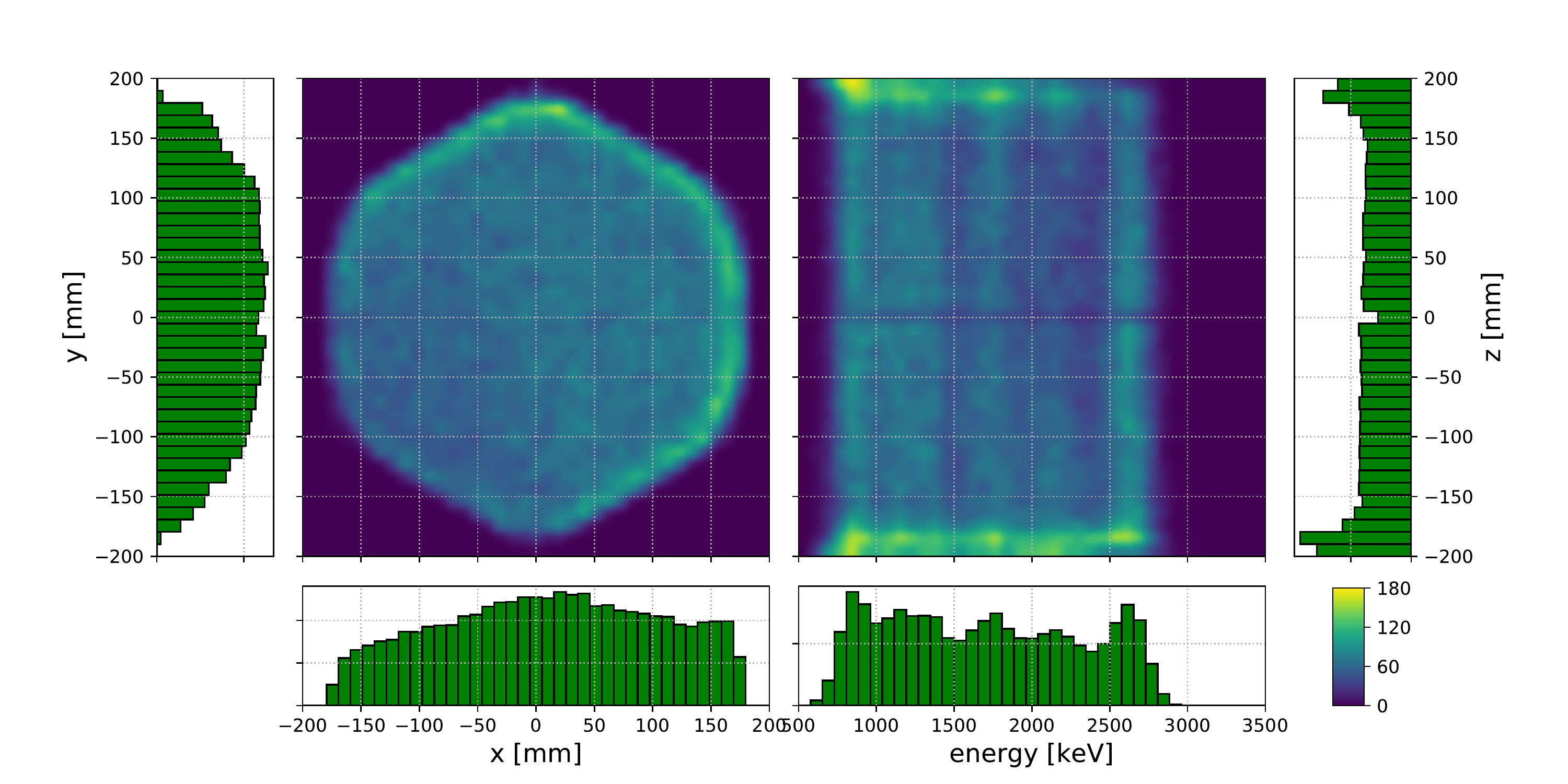}
\caption{Event distribution of the training set. The distinct locations of the calibration sources outside of the TPC causes the detector volume to be illuminated non-uniformly. The process of discarding some events based on the original event distribution (see text) creates a more uniform distribution in all spatial dimensions and energy.}
\label{fig:ScintRec_TrainingSet_EventDistribution}
\end{figure}

This image is then fed to a convolutional neural network, which is described in detail below. The dataset used for this study consists of source calibration runs. During a calibration run, one of the four available radioactive sources ($^{228}$Th, $^{226}$Ra, $^{60}$Co, $^{137}$Cs) is deployed to specific, well known locations outside of the TPC. Due to the relatively short interaction length of gamma rays in LXe, the event distribution is strongly biased towards the location of the source. We combine several runs from different source positions, taken over a period of \num{6} months, to acquire a sufficiently large training set. We use the event information extracted by the conventional \mbox{EXO-200} reconstruction to apply event selection cuts -- namely the number of scintillation clusters, the event multiplicity and charge energy -- and retrieve the true labels. Note that this information could also be determined by means of DNNs such as the ones described above. However, in this study we do not demonstrate this linkage due to concurrent development of both DNN algorithms. We select events where only a single energy deposit (SS) occurred and only one prompt scintillation signal was found. We apply an energy threshold of \SI{800}{\kilo\electronvolt}, as measured by charge, and cut events with high scintillation and low charge yield to exclude the detection of $\alpha$-particles, which exhibit a signal pattern different from high energy electrons that occur after $\gamma$-detections or $\beta$-decays in the TPC. The remaining set of events is further reduced in order to create a data set with uniform position and energy distribution. This step turned out to be necessary in order to improve the model's ability to generalize on new, unseen data, in particular for event positions that occur at low frequency in the training data. From the original event distribution with $P=(x,y,z,E)$ we build a \num{4} dimensional probability density function (PDF) using a Gaussian Kernel Density Estimation (KDE) \cite{Parzen:1962}. We then loop through all events, draw a random number $\rho\in\mathcal{U}(0,1)$ from a uniform distribution, and keep the event if
\begin{equation}
\rho<\kappa\cdot\frac{PDF(P_{\mathrm{min}})}{PDF(P)}
\end{equation}
where $P_{\mathrm{min}}$ is location of lowest statistics and $\kappa$ a margin factor allowing to trade off some uniformity for an increased number of events. Figure~\ref{fig:ScintRec_TrainingSet_EventDistribution} shows the distribution of events in the training set after applying the described procedure. In total, \num{\sim 70000} events are kept, from which \num{1024} events (\SI{1.5}{\percent}) are randomly selected and split into a validation set. The events which are removed by the described procedure are stored in separate files and can be used to evaluate the trained model.

\subsection{Network architecture}
\begin{figure}
\centering
\includegraphics[width=\textwidth]{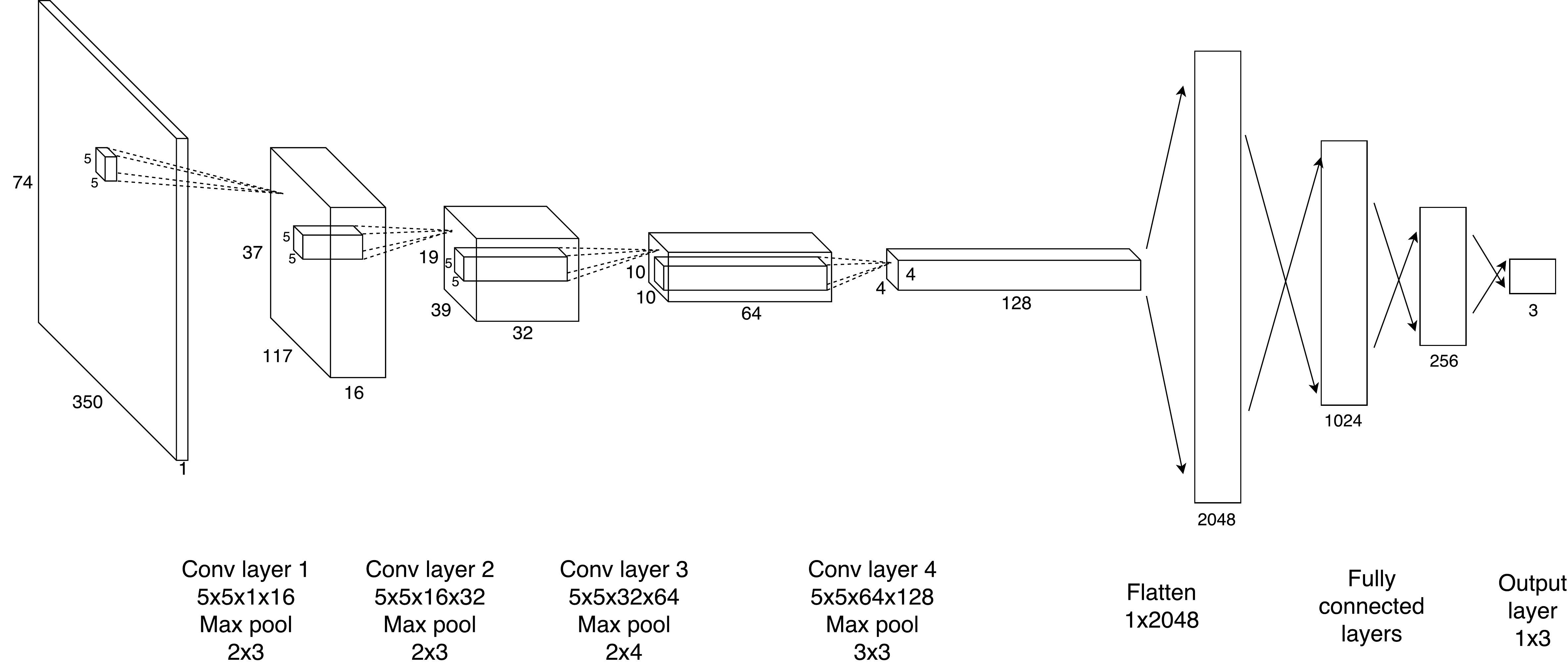}
\caption{Architecture of the deep neural network for the scintillation light reconstruction. Four convolutional layers with increasing number of kernels and constant kernel size are followed by three densely connected layers with a final output layer consisting of three units.}
\label{fig:ScintRec_conv_net_architecture}
\end{figure}

\begin{figure}
\centering
\includegraphics[width=0.8\textwidth]{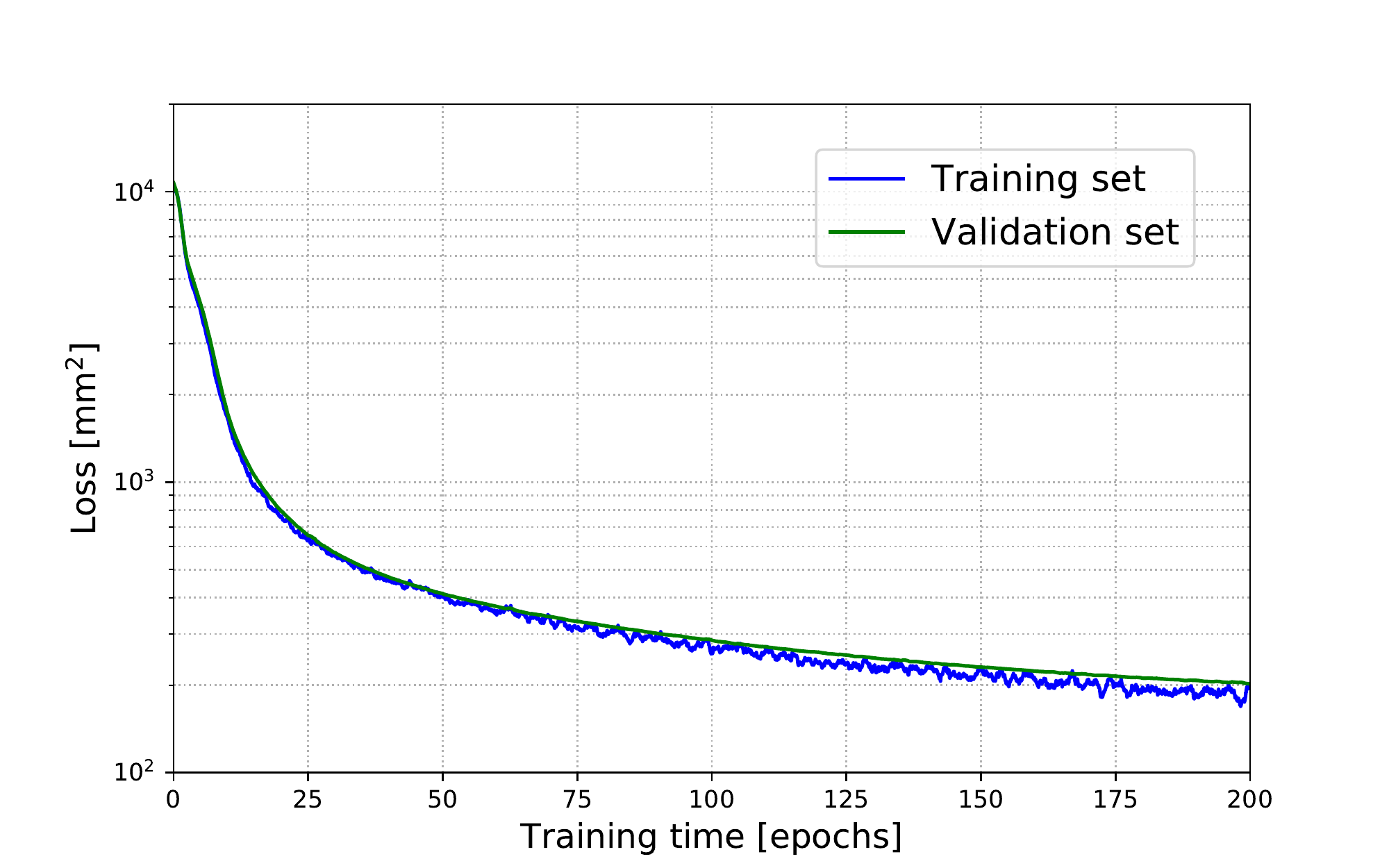}
\caption{Training and validation loss as a function of training epochs.}
\label{fig:ScintRec_loss}
\end{figure}

For this study we construct a convolutional neural network that takes event images like the one illustrated in Figure~\ref{fig:ScintRec_Image} (left) as input and has three units in the output layer corresponding to the x-,y- and z-position of the event. Given the choice of selecting \num{350} samples per waveform and \num{74} waveforms in total containing the unique signature of the event position, a convolutional neural network seems most suitable to handle the \num{25900} input features. The structure of the neural network is illustrated in Figure~\ref{fig:ScintRec_conv_net_architecture}. The first convolutional layer has \num{16} kernels of size $\num{5}\times\num{5}$ with a stride of \num{1} followed by a maximum pooling layer of size $\num{2}$(channel wise)$\times\num{3}$(time-wise). This pattern is repeated in layer \num{2} while increasing the number of kernels to \num{32}. In the following two layers, the number of kernels with the same size is increased to \num{64} and \num{128} respectively with max pooling layers of size $\num{2}$(channel-wise)$\times\num{4}$(time-wise) and $\num{3}$(channel-wise)$\times\num{3}$(time-wise) in between. After the last pooling layer, the structure is transformed into a one-dimensional array, yielding \num{2048} units, which are fed to a densely connected network reducing the size of units per layer to \num{1024}, \num{256} and finally \num{3} in the output layer.
It consists of \num{\sim 2.4e6} trainable parameters which are initialized using Glorot initializer. We use an Adam optimizer to minimize the loss function $L = C + \lambda\cdot R$, where 
\begin{equation}
C = \frac{1}{3m}\sum_{i=1}^{m}\sum_{k=1}^{3}\left(y_{i}^{k}-\hat{y}_{i}^{k}\right)^{2}
\end{equation}
is the mean square deviation of the predicted and true position over all coordinates with $m$ being the size of the mini batch. $\lambda\cdot R$ is the $L$2 regularization term where $R$ is defined in Equation~\ref{eq:U-reg} adapted to the architecture described here. We have tuned the hyper-parameters such as the size of the neural network, learning rate ($\alpha=\num{2e-6}$) and regularization ($\lambda=\num{5e-3}$) for best performance by testing the model during training on the validation dataset. After training the network for \num{200} epochs, the improvement in the loss is \SI{< 0.6}{\milli\meter\squared\per epoch} while the variance is sufficiently small as shown in Figure~\ref{fig:ScintRec_loss}.
The remaining bias does not improve with increased size of the network and can therefore be attributed to the finite position resolution and limited amount of events in the training data set. The loss may be further decreased by adding more data to the training set, but this is not demonstrated in this study. The value of the loss reaches \SI{200}{\milli\meter\squared} after \num{200} epochs, which translates to an accuracy of $\sigma_{3D}$~=~\SI{24.5}{\milli\meter}. We deem training to be complete at this point and evaluate the accuracy of the model on an independent set of data.

\subsection{Testing of the trained DNN}
\begin{figure}[htp]
\centering
\subfloat[$^{228}$Th at S5]{\includegraphics[width=\textwidth]{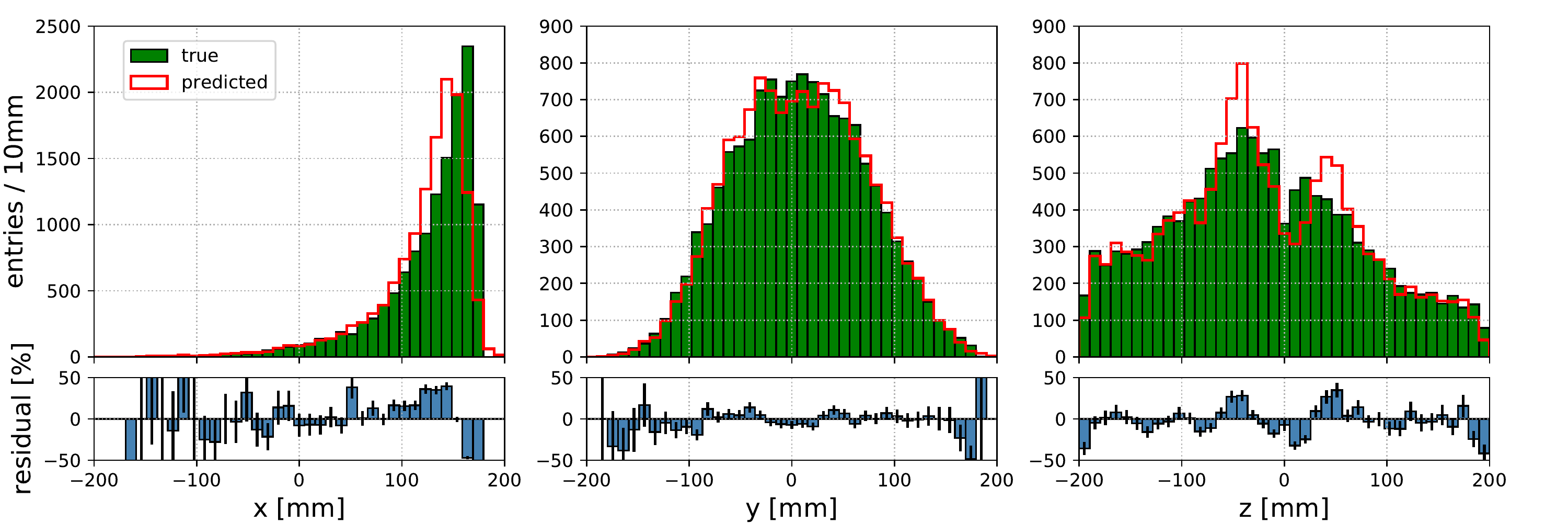}} \\
\subfloat[$^{228}$Th at S11]{\includegraphics[width=\textwidth]{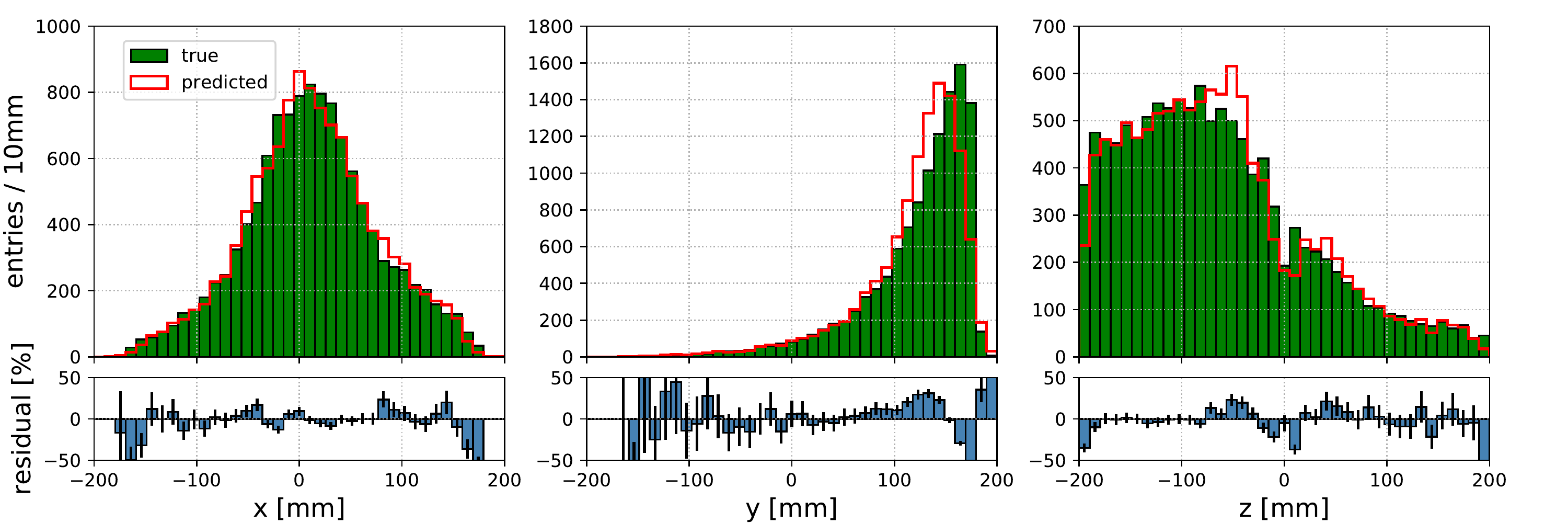}}
\caption{Evaluation of the trained model on an independent set of source data. Events removed from the training set during the process of generating a uniform event distribution (see text) are used to compare the predicted position by the model to the true position from charge information for a $^{228}$Th run at S5 (a) and S11 (b).}
\label{fig:ScintRec_EvalTh}
\end{figure}

\begin{figure}[htp]
\centering
\includegraphics[width=\textwidth]{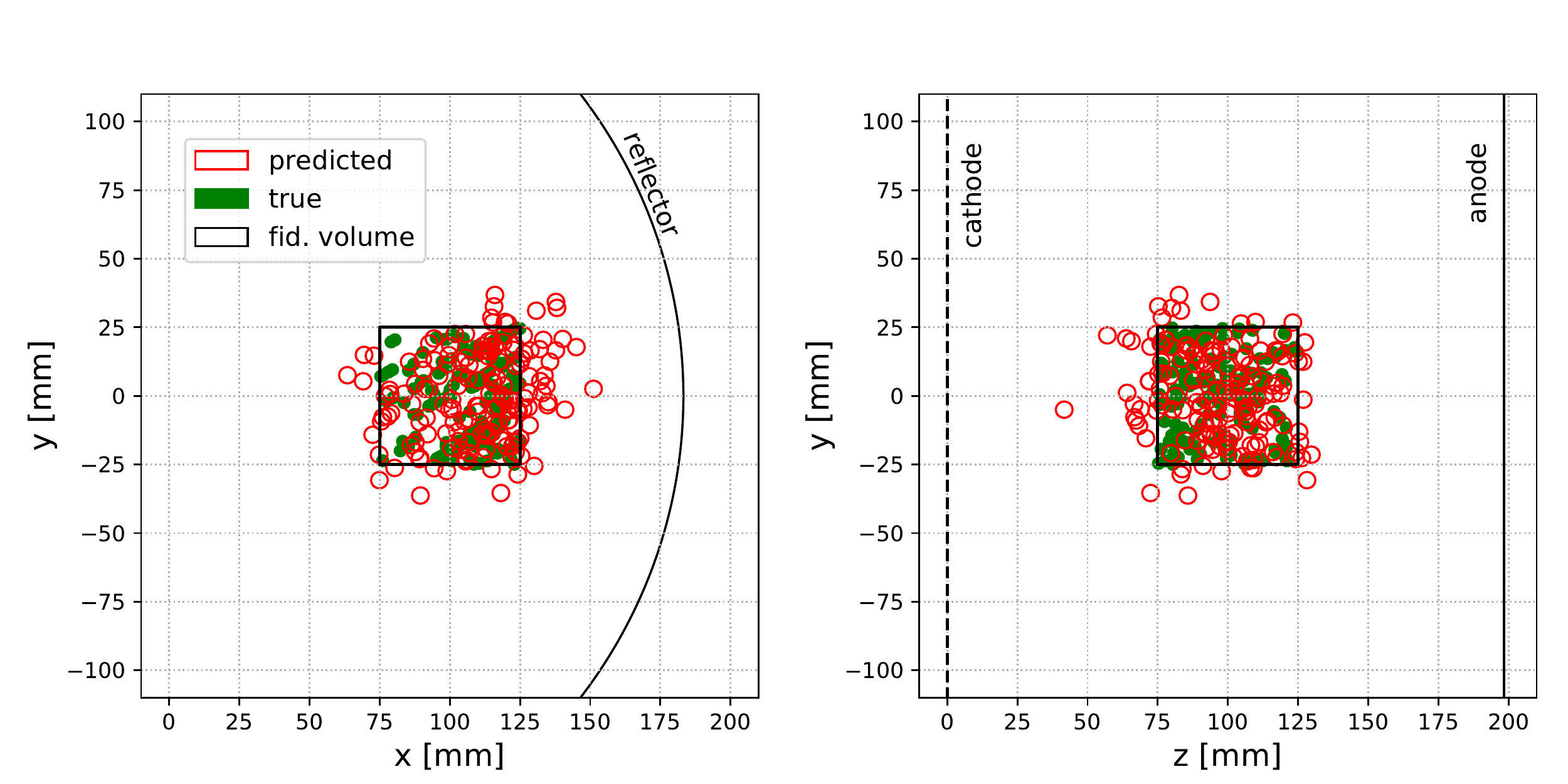}
\caption{Evaluation of the trained model on an independent set of test data. The events are selected to be in a tight fiducial volume of $\SI{50}{\milli\meter}\times\SI{50}{\milli\meter}\times\SI{50}{\milli\meter}$ at position $x=\SI{100}{\milli\meter}$, $y=\SI{0}{\milli\meter}$ and $z=\SI{100}{\milli\meter}$ and have energies above \SI{2400}{\kilo\electronvolt}.}
\label{fig:ScintRec_Eval_fv1}
\end{figure}

The accuracy of the trained DNN is tested using various sets of independent data. We utilize the events that were discarded during the process of creating a training set with uniform event distribution. Each test set contains the events of only one source run with specific source type and location. Therefore, the event distribution is biased towards the location of the source and differs from that of the training set. Figure~\ref{fig:ScintRec_EvalTh}~(a) shows the result of evaluating the model on a $^{228}$Th source run at position S5 separated into x-, y- and z-coordinates. The green distributions correspond to the true position as determined by charge information whereas the red distributions show the predicted position from the scintillation signal.
The bottom panels show the residuals defined as (predicted~-~true)/true. The total number of tested events is 13211 and the averaged distance between true and predicted event position in 3D space is $\overline{d}=\SI{22.5}{\milli\meter}$. The averaged distances in x-, y- and z-coordinates separately are: $\overline{d_{x}}=\SI{13.6}{\milli\meter}$, $\overline{d_{y}}=\SI{11.3}{\milli\meter}$, $\overline{d_{z}}=\SI{8.1}{\milli\meter}$. We have also evaluated the model on different source types, including $^{226}$Ra and $^{60}$Co, at various source positions and similar accuracies for the predicted event positions were determined. The result for a $^{228}$Th run at S11 is shown in Figure~\ref{fig:ScintRec_EvalTh}~(b). In this case, the averaged distance between the predicted and true event position is \SI{22.4}{\milli\meter}, while $\overline{d_{x}}=\SI{11.8}{\milli\meter}$, $\overline{d_{y}}=\SI{12.7}{\milli\meter}$ and $\overline{d_{z}}=\SI{8.6}{\milli\meter}$. For both $^{228}$Th runs at S5 and S11 the discrepancy between predicted and true position is largest in the coordinate for which the event distribution is close to detector edges and the Teflon reflector (x-coordinate for S5 and y-coordinate for S11).
\begin{figure}[htp]
\centering
\subfloat[$^{228}$Th at S5]{\includegraphics[width=\textwidth]{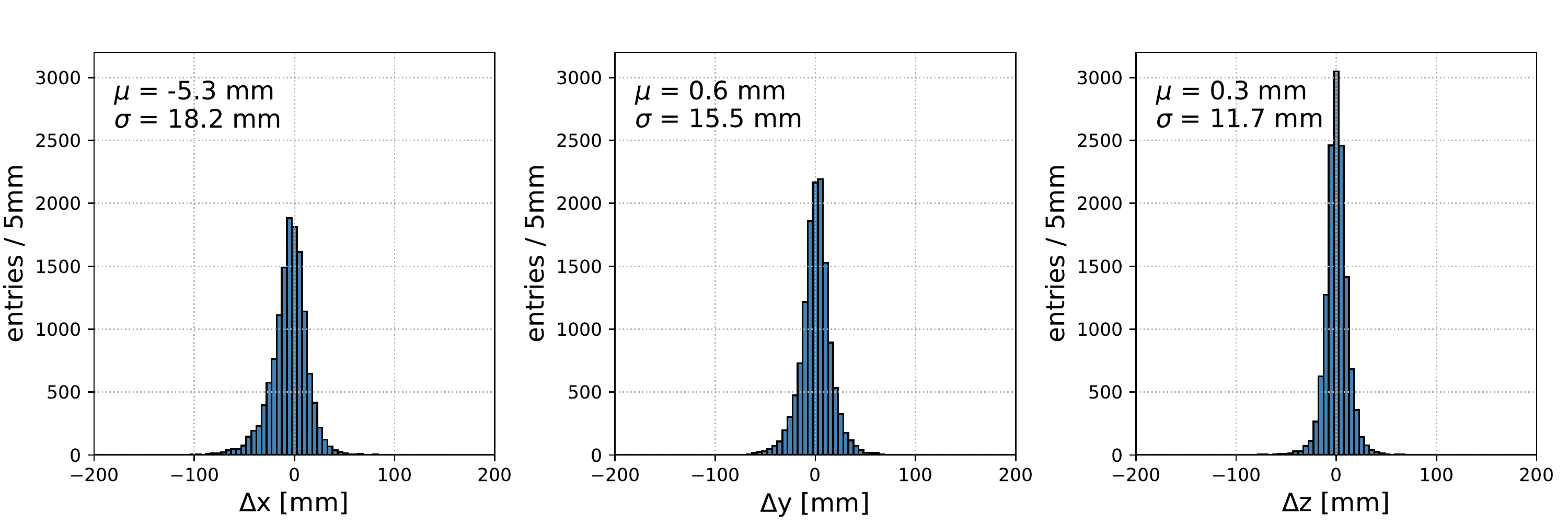}} \\
\subfloat[$^{228}$Th at S11]{\includegraphics[width=\textwidth]{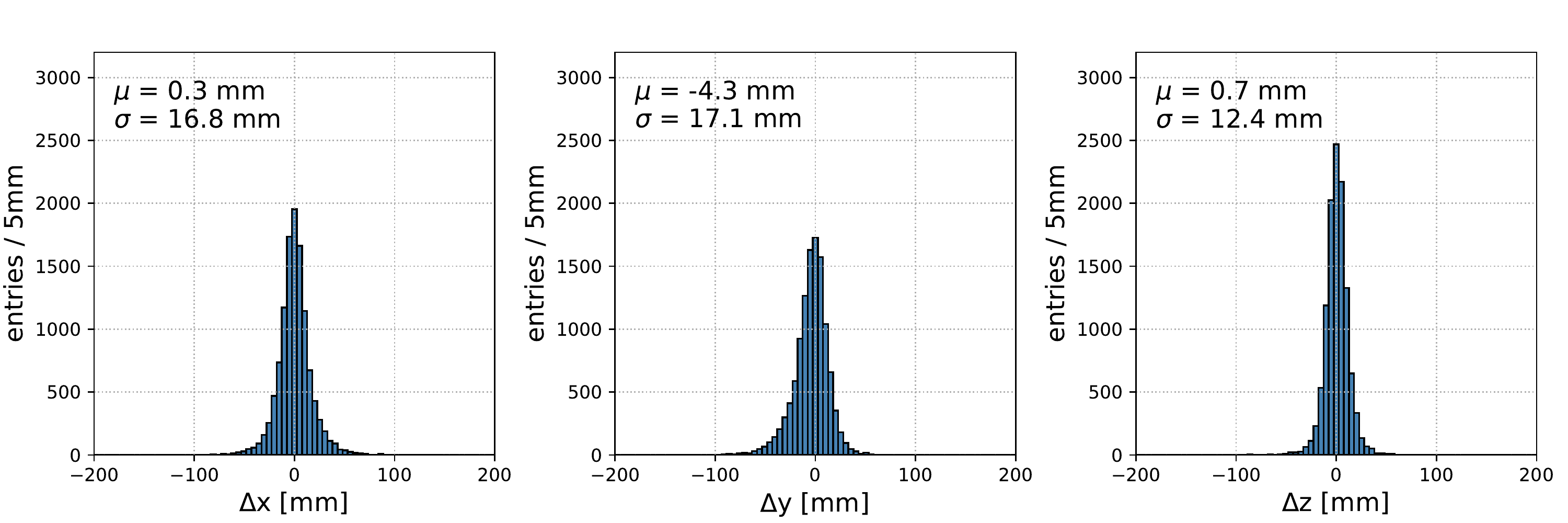}}
\caption{Distribution of difference in predicted and true event position for a $^{228}$Th run at S5 (a) and S11 (b). The non-zero mean in X-coordinate for S5 and Y-coordinate for S11 indicate the larger discrepancy when the event distribution is shifted towards the edge of the detector and close to the Teflon reflector.}
\label{fig:ScintRec_EvalTh_posDiff}
\end{figure}
This discrepancy is further illustrated in Figure~\ref{fig:ScintRec_EvalTh_posDiff} which shows the distribution of (predicted~-~true) event positions where the mean is non-zero in these cases.  However, this shift is several times smaller than the standard deviation of these distributions. Possible explanations are the fact that events occurring near the Teflon reflector might get perturbed due to inhomogeneity in the drift field and charge loss. In such cases the true position deviates from the information retrieved from charge collection. Moreover, the proximity to the Teflon reflector causes more light to be reflected before detection by the APDs which complicates the reconstruction of the light distribution. In addition to the individual source runs, we have generated a test set consisting of events which are contained in a tight fiducial volume of $\SI{50}{\milli\meter}\times\SI{50}{\milli\meter}\times\SI{50}{\milli\meter}$ at position $x=\SI{100}{\milli\meter}$, $y=\SI{0}{\milli\meter}$ and $z=\SI{100}{\milli\meter}$ with energies above \SI{2400}{\kilo\electronvolt}. The event distribution of this data set is manually constructed and hence is completely independent of the training data. The data set contains \num{160} events. The true positions of these events are shown as green dots in Figure~\ref{fig:ScintRec_Eval_fv1} while the predicted positions are represented as red circles. 
A very good agreement between true and predicted position is visible. The averaged distance is \SI{13.0}{\milli\meter}, while the contributions from each coordinate are: $\overline{d_{x}}=\SI{8.3}{\milli\meter}$, $\overline{d_{y}}=\SI{4.8}{\milli\meter}$, $\overline{d_{z}}=\SI{6.2}{\milli\meter}$. The smaller averaged distances as compared to the previous test sets is mainly attributed to the guaranteed large signal to noise ratio, due to the energy cut, and the distance of the selected events to the detector edges as described previously.

\section{Conclusions}
This work presents the first successful reconstruction of energy and position of events in \mbox{EXO-200} with deep learning. We found that the energy resolution, as measured for a calibration gamma source, is slightly better than that of the conventional reconstruction. 

Other works already pursued the reduction of reliance on conventional reconstruction and its possible imperfections, with Ref.~\cite{microboone:2017} using only the step of noise filtering before waveforms are used as input for a DNN. We continue this trend and push it a step forward by using, essentially, raw waveforms. Raw waveforms were used in this study to directly extract the energy and position of an event without applying any higher level routines, such as noise filtering.

More importantly, the cited DNN approaches have only been validated with MC. This means that they may not hold their promise when faced with real data. This work has demonstrated that this risk is real. We found a way to mitigate this risk in a particular case of \mbox{EXO-200} charge energy reconstruction.

We validate the performance of reconstruction of charge energy and position of events with real calibration data. To our knowledge, this is the first time that the real detector data is used in validation of a DNN algorithm in high energy and neutrino physics. We believe that data driven validation is a necessary prerequisite for acceptance of any DNN based reconstruction. 

The position of events in \mbox{EXO-200} was also successfully reconstructed using raw APD waveforms that did not undergo the denoising procedure. Due to the absence of a detailed MC simulation of light propagation and collection in \mbox{EXO-200}, we used the event position obtained from U- and V-Wire signals to provide the truth input during the DNN training. The validation of the position reconstruction was also done with the calibration data. 

The full \mbox{EXO-200} dataset spans several years. In order to use the approaches developed here on the full dataset one needs to also apply corrections for time-dependent effects, e.g., for variation of the finite electron lifetime in the detector. The methods to derive such corrections rely on the same conventional reconstruction routines and are not believed to pose a substantially different challenge to the DNN approach. Nevertheless, the effort of determining the time dependent corrections with DNNs has not been attempted. So this work does not demonstrate a successful reconstruction of the full \mbox{EXO-200} dataset, only of a set of randomly chosen calibration data runs. Reconstructing the full \mbox{EXO-200} dataset; performing higher level analysis of \mbox{EXO-200}, such as event identification and classification; and further reduction of the reliance of the DNNs on MC during training are the natural avenues for future work.

Lastly, in addition to the performance advantages of the DNN approaches discussed here and in the referenced publications, one can also notice the following. The DNN approaches described here required surprisingly little time to set up and begin to perform competitively. Partly, this is because the DNN approaches take advantage of already existing free but professionally developed and maintained software libraries, like TensorFlow and Keras. With continued scrutiny and advances, it may become possible that the DNN based reconstruction and data analysis will be employed as first choice of the next generation experiments in neutrino, dark matter, and high energy physics.

\acknowledgments
We gratefully acknowledge the support of Nvidia Corporation with the donation of the Titan Xp GPU used for this research. We thank Adam Coates (Baidu) for helpful and encouraging discussions and Evan Racah (NERSC, LBNL) for support of deep learning applications at NERSC at the beginning of this work. We thank the Erlangen Regional Computing Center (RRZE) for the compute resources and support. We thank the Deutsche Forschungsgemeinschaft (DFG) for the support of this study. \mbox{EXO-200} data analysis and simulation uses resources of the National Energy Research Scientific Computing Center (NERSC), which is supported by the Office of Science of the U.S. Department of Energy under Contract No. DE-AC02-05CH11231. EXO-200 is supported by DOE and NSF in the U.S., NSERC in Canada, SNF in Switzerland, IBS in Korea, RFBR in Russia, DFG in Germany, and CAS and ISTCP in China. We gratefully acknowledge the KARMEN Collaboration for supplying the cosmic-ray veto detectors, and the WIPP for their hospitality.

\bibliographystyle{JHEP} 
\bibliography{references}

\end{document}